# Three-dimensional imaging of integrated-circuit activity using quantum defects in diamond

Marwa Garsi[1,2]*, Rainer Stöhr[1], Andrej Denisenko[2], Farida Shagieva[2], Nils Trautmann[3], Ulrich Vogl[3], Badou Sene[4], Florian Kaiser[1†], Andrea Zappe[1], Rolf Reuter[1], Jörg Wrachtrup[1]

**Affiliations**

1) 3rd Institute of Physics, IQST, and Research Center SCoPE, University of Stuttgart, 70569 Stuttgart, Germany.
2) SQUTEC-TTI GmbH, 70569 Stuttgart, Germany.
3) Corporate Research and Technology, Carl Zeiss AG, Carl-Zeiss-Strasse 22, 73447 Oberkochen, Germany.
4) Mobility Electronics, Robert Bosch GmbH, 72762 Reutlingen, Germany.

† Now at Luxembourg Institute of Science and Technology (LIST)

* m.garsi@pi3.uni-stuttgart.de

**Abstract**

The continuous scaling of semiconductor-based technologies to micron and sub-micron regimes has resulted in higher device density and lower power dissipation. Many physical phenomena such as self-heating or current leakage become significant at such scales, and mapping current densities to reveal these features is decisive for the development of modern electronics. However, advanced non-invasive technologies either offer low sensitivity or poor spatial resolution and are limited to two-dimensional spatial mapping. Here we use near-surface nitrogen-vacancy centres in diamond to probe Oersted fields created by current flowing within a multi-layered integrated circuit in pre-development. We show the reconstruction of the three-dimensional components of the current density with a magnitude down to about $\approx 10\ \mu A/\mu m^2$ and sub-micron spatial resolution at room temperature. We also report the localisation of currents in different layers and observe anomalous current flow in an electronic chip. Our method provides, therefore a decisive step toward three-dimensional current mapping in technologically relevant nanoscale electronics chips.

## I. Introduction

The rapid growth and downscaling of silicon integrated circuits (ICs) have ushered revolutions in many areas of today's society [1–4], such as high-speed internet [5], in-car navigation [6] and leadless pacemakers [7]. However, if the semiconductor community has underpinned Moore's law [8] for over 50 years by shrinking the size of electronic components, the scaling roadmap is nearing its end [1,9]. As a result, next-generation technologies like autonomous driving [10] or quantum processors [11] rely on a new strategy: three-dimensional chip architectures [12–16]. In this regard, device development, optimisation and failure analysis are severely challenged due to the absence of methods for direct visualisation of three-dimensional charge flow. This particularly concerns multi-layer chips with sub-micron feature sizes.

Most electric current imaging techniques visualise charge transport through the associated magnetic fields that pass unaffected through the materials used in semiconductor devices. One approach consists of delayering the chip to probe fields with a micro-needle [17]. Although this technique enables high spatial resolution, it inherently alters the current path. Non-destructive current imaging can be implemented using superconducting quantum interference device (SQUID) microscopes, but the inherent stand-off distance limits the spatial resolution to tens of micrometres [18]. Alternatively, giant magneto-resistance (GMR) microscopes provide excellent spatial resolution but come at the expense of much lower field sensitivities [19,20]. However, SQUID and GMR microscopes are only sensitive to a single magnetic field component, limiting reliable current imaging to the two-dimensional realm.

In this article, we demonstrate current imaging in a three-dimensional integrated circuit using quantum sensors at room temperature. We use nanoscale nitrogen-vacancy (NV) centres in diamond [21,22] which offer the unique property to probe all three vectorial components of a magnetic field simultaneously on the nanoscale [23,24] and in a non-invasive fashion. Besides, NV centres operate under a wide span of external conditions [25–29] and demonstrate excellent sensitivity to magnetic fields [30,31].

Pioneering work successfully demonstrated current imaging on the nanoscale [32,33] and IC activity imaging [34–36] with NV centres but has been so far restricted to two-dimensional current imaging. In this work, we demonstrate three-dimensional current distribution imaging within a micro-chip designed with the recent back end of line (BEOL) technology [13,37]. For this, we employ an NV-based wide-field microscope described in Fig. 1a, to synchronously map vectorial magnetic fields over a region of 90 µm x 90 µm. We use the instrument to measure the current density flow in the multi-layered IC (Fig. 1b), notably without using prior knowledge about its design during the analysis process.

## II. Current density imaging using NV centres

The principle of the experiment is depicted in Fig. 1c. Long-range magnetic fields, also known as Oersted fields, are created by moving charges according to the Biot-Savart law given by:

$$\boldsymbol{B}(\boldsymbol{r}) = \frac{\mu_0}{4\pi} \iiint \frac{\boldsymbol{J}(\boldsymbol{r}') \times (\boldsymbol{r} - \boldsymbol{r}')}{|\boldsymbol{r} - \boldsymbol{r}'|^3} d^3r, \quad (1)$$

where $\mu_0$ is the vacuum permeability, $\boldsymbol{r}$ represents the spatial coordinates at the observation point, and $\boldsymbol{J}(\boldsymbol{r}')$ is the current distribution in the source plane.

The magnetic field isolines in Fig. 1c show that magnetic field contributions merge with distance from the current source, resulting in blurry patterns. In our experiment, we place a diamond homogeneously implanted with near-surface NV centres in the vicinity of the current flow, at only a few hundred nanometres from the surface of the IC. The electron spin of each NV centre is thus affected by the magnetic field via the Zeeman interaction $\mathcal{H}_{\mathrm{EZ}} = -\gamma_{\mathrm{NV}} B \cdot S$ where $\gamma_{\mathrm{NV}}$ is the NV-associated electron spin gyromagnetic ratio, $\boldsymbol{B}$ is the total magnetic field in the vicinity of the NV centre and **S** represents the spin operators for the electron spin with S = 1.

Probing this Zeeman interaction on the multiple NV orientations, naturally occurring in the diamond lattice (Fig. 1d), is done by performing optically detected magnetic resonance (ODMR) on the NV centres [24] (see Supplemental Material [38]). We inject a total current $I$ = 19.8 mA into the circuit which splits into several sub-paths, creating distinct local Zeeman shifts on the ODMR spectra (Fig. 1e). For each pixel of our image, we fit the spectrum to extract the eight resonance frequencies. Finally, we compare the extracted resonance frequencies to the ground-state NV spin Hamiltonian, including the zero-field splitting, the Zeeman and the Stark effects (see Supplemental Material [38]).

We perform the experiment on two distinct chips labelled as device 1 and device 2. Investigation of both samples under light-microscopy (Fig. 2, a and b) reveals no difference. On the contrary, mapping the Oersted fields exposes a failure immediately. With device 1, Fig. 2c shows Oesterd fields clearly reflecting the geometry of the underlying structure. In Fig. 2d, we can see that device 2 produces nearly one order of magnitude lower magnetic fields (a maximum amplitude of $|B_x|$ = 513(6) µT compared to 73(5) µT). Furthermore, the magnetic field patterns $B_y$ and $B_z$ produced by device 2 mismatch with those produced by device 1. Thus, already comparing the magnetic field produced by the two devices can help identifying anomalous behaviour given the high dynamic range of the NV centres.

To better understand the current distribution producing such Oersted field patterns, we reconstruct the lateral current density $J_{xy}$. We follow the procedure described in references [39,40] and use the components of the magnetic field $B_x$ and $B_y$ to numerically invert the Biot-Savart law (Eq. (1)), resulting in the lateral current density $|J_{xy}|$ shown in Fig. 2, e and f (see Supplemental Material [38]). It is important to note that since the magnetic field signals are collected outside the device to perform a non-destructive measurement, all the fields generated in different layers merge, resulting in a flattened image. This characteristic makes non-destructive imaging of three-dimensional ICs difficult, and thus, testing and validation of modern chips challenging [41]. In Fig. 2e, the current paths in device 1 follow the shape of the visible structure shown in Fig. 2a. A closer look at the central part of the map reveals a weak current contribution with wide lateral spreading, indicating that additional currents flow underneath. Finally, the flow appears weaker in some parts of the circuit, like at the sharp corners. We can see in Fig. 2f that several current sources produce fields of similar intensity in device 2. Observing $|J_{xy}|$ alone, i.e. without further information about the field distribution over the vertical axis $z$, is insufficient to comprehend the current path.

In section III, we investigate the different layer contributions to locate the flow within the device and in section IV, we seek the third dimension of the current density, $J_z$.

## III. Localisation of currents inside a multi-layered device

To resolve the signal in the vertical direction $z$, we investigate different linecuts along $x$ in the magnetic field map $B_x$ (Fig. 3). We fit the linecuts with the Biot-Savart model (Eq. (1)), using the infinite wire approximation (Eq. (2)),

$$B_{x,y} = \frac{\mu_0 I_{y,x} \Delta z}{2\pi \left[ \left| r_{xy} - r_{\text{wire}} \right|^2 + \Delta z^2 \right]} + o \qquad (2)$$

where $I_{y,x}$ is the lateral current amplitude, $r_{xy}$ represents the observation position on the $xy$-plane, $r_{\text{wire}}$ the position of the current source on the $xy$-plane, Δz is the distance between the current source and the observation position on the vertical axis $z$ and $o$ is a constant offset (Fig. 3a).

The fitting procedure reveals a contribution from two layers: the first one at $\Delta z_1 = 4.5(5)$ µm away from the layer of NV centres and the second one at $\Delta z_2 = 8.5(8)$ µm (see Supplemental Material [38]). These distances are validated by comparison with the actual values, estimated to $\Delta z_{1,\text{ true}} = 4.5(1)$ µm and $\Delta z_{2,\text{ true}} = 7.9(1)$ µm. For device 1 (device 2), we identify a current of amplitude $I_A$ ($I_{A'}$) in the main lead, dividing into currents of amplitude $I_{B1}$ ($I_{B'1}$) and $I_{B3}$ ($I_{B'3}$) in the split branches (Fig. 3, b and c). By referring to the expected values of direct current in the main lead and the split branches given by the manufacturer, we can conclude that the device 2 presents a failure. When comparing the results from device 2 (defective) to device 1 (operating), most of the loss appears on the outer layer (at $\Delta z_1$), presenting one order of magnitude lower current amplitude. In contrast, the deeper layer (at $\Delta z_2$) presents a smaller loss. Finally, the analysis of other line profiles reveals another current contribution at $\Delta z_2$ present in the operating and defective devices, in both cases with no apparent anomaly (see Supplemental Material [38]). From these observations, we conclude that failure happens in the layer at $\Delta z_2$ and then affects the outer layer by propagation.

Overall, the simple model with infinite wire approximation already shows excellent agreement with the experimental data. In order to verify the consistency of the procedure, we now perform a simulation of Oersted fields produced by a multi-layered device. The simulation reproduces the layering of the chip, made of a SiGe technology described in [37], the sensors-device geometry, and some of the apparent geometric features of the chip for guidance only.

The total thickness of the simulated structure is 11.8 µm and combines twelve stacked layers shown in Fig. 4a. As depicted in Fig. 4b, two layers across the structure are electrically active and labelled as 1st and 2nd active layer (AL). Through-silicon vias (TSVs) connect the 1st AL to the bottom layer of the structure. We investigate magnetic fields generated by this structure, resulting in patterns at the position of the sensors shown in Fig. 4c. Similarly to the experimental observations (Fig. 2), the contribution from the 1st AL is clearly defined and unambiguously related to the shape of the structure. The contribution from the 2nd AL shows a pronounced lateral spreading, and the signal arising from two distinct wires starts to blur out. Finally, the contribution from the vertical current is weak due to both the observation position and the presence of counter-propagating flows which average out magnetic field contributions (see Supplemental

Material [38]). Still, a current propagating vertically has a nonzero contribution in $B_{xy}$ compared to its contribution in $B_z$ (see Supplemental Material [38]). Therefore, currents propagating in the *z* direction can be sensed by NV centres contrary to magnetometers such as SQUID and GMR sensors since they only measure the out-of-plane component of the magnetic field $B_z$. These magnetometers indirectly identify vertical currents by monitoring a discontinuity in the current path. However, observing the current discontinuity can be difficult to interpret since signals produced by counter-propagative sources can cancel each other. Besides, as further discussed in the Supplemental Material [38], analysing signals from multi-layered devices is more reliable with NV centres since there is no ambiguity in identifying orthogonal overlapping signals.

Lastly, to study the flow in the three-dimensional structure and observe vertical currents, we can infer information about the third component of the current density, $J_z$.

## IV. Three-dimensional current density mapping

The current-carrying wires have a non-negligible thickness of a few hundred nanometres, leading to a possible contribution of the current's *z*-component. In order to evaluate the total current density in all directions, we now consider a component $J_z \neq 0$ in Eq. (1) (see Supplemental Material [38]). The resulting maps are depicted in Fig. 5a-c and the results suggest three-dimensional contribution of the current flow inside the wires. It is important to note that the maps shown Fig. 5a-c, still represent a signal where all the contributions over the vertical axis *z* are merged. Using the fitting algorithm employed in Fig. 3 over the entire map would enable the selection of the signal from each layer separately. However, the layout must be known for devices thick of several microns to reconstruct the current density reliably over the entire chip since the magnetic field spreading from the deep layers is too prominent. Thus, in Fig. 5d, we only represent the current in the outer lead for visualisation. We can see in Fig. 5, c and d, a non-negligible current flow in the *z*-direction at the edges and the corners of the leads. Thus, using a two-dimensional model for a three-dimensional device can lead to locally underestimating the current amplitude. For example, a weaker current density was observed at the corners of the split branches when considering a lateral flow only (Fig. 2). This information can be crucial for evaluating current crowding at corners in interconnect structures, which plays an essential role in nucleating voids and hence failure of ICs [42–44]. More importantly, having access to the full-vector information of the current density helps quantify and understand the current flow through different stacks in layered materials. For instance, in the outer layer, we can observe a prominent $J_z$ contribution at the edge of the main lead. As for the simulation (Fig. 4), this contribution is the averaged result from counter-propagating currents in vias. Current in the main lead flows down to a deeper layer where it splits into two paths and goes back to the outer layer to further flow in the split branches (see Supplemental Material [38]). As the component $B_z$ does not carry information about $J_z$, and $B_{xy}$ shows a specific pattern with the presence of counter-propagating fields, developing an algorithm using $\left(B_{xy} - B_z\right)$ and pattern recognition techniques [45] is the next step to identify the contribution from each current source over the entire device.

## V. Conclusion

Using NV centres in diamond, we have demonstrated imaging of three-dimensional current density in a multi-layered integrated circuit. First, we compared the current flow in two devices, primarily identical under light microscopy, and identified an operational chip and a defective one. Exploiting the NV centre's high dynamic range, we observed one order of magnitude lower current amplitude in the defective device, which demonstrate its potential use for failure analysis of ICs. Furthermore, we have shown how to localise currents originating in different leads even with multiple stacked layers. Lastly, we have presented imaging of the three vectorial components of the current density. Although the out-of-plane component of the current density $J_z$ is generally neglected in current density imaging techniques, we revealed a significant out-of-plane contribution of $J_z$ close to sharp edges. Finally, we have discussed how to image three-dimensional current density over the entire device. In this regard, resolving currents from different sources across the structure depends on the spatial resolution of the imaging technique. NV centres offer the closest sensor-sample proximity known so far and can demonstrate spatial resolution of only a few tens of nanometres with a scanning probe setup [32]. However, the device's configuration, including the capping layer, limits the spatial resolution of the magnetic fields and, thus, of the current density maps. In order to non-destructively resolve each layer with higher resolution, one solution is to interpolate the current distribution at the source plane using additional layout information, which can be obtained directly using circuit designs or ptychographic X-ray laminography techniques [46].

Additionally, NV centres can be used to explore various magnetic field regimes [47–51] and can be operated with high-speed imaging [52,53]. Combining a practical NV-based imager [54] with X-ray imaging [55] will provide complete information on nanoscale three-dimensional current-carrying structures [56]. Such an imager is thus particularly relevant for designing modern three-dimensional electronic chips where failure analysis is primordial to predict and determine the root cause of a malfunction and thus steer the manufacturing process adequately at an early stage [41].

Finally, unravelling three-dimensional electronic signals using NV centres will leverage further advancements in many areas. For instance, it will serve neuroimaging to overcome the limits of conventional current density imaging techniques and help to reveal new features [57]. The NV-based microscope would serve to observe charge transport in diverse multi-layered electronic systems in science and technology [58,59].

## Appendix A. Diamond substrate

The diamond plates are fabricated by combining electron-beam lithography and reactive ion plasma etching. Prior to fabrication, NV centres are created by implantation of nitrogen with an energy of 9.8 keV and an implantation dose of $2 \times 10^{12}$ 1 cm$^2$ into a 100-oriented CVD diamond (electronic grade, Element Six). After implantation, the diamond is annealed at 960°C for 2 hours at a pressure of $10^{-7}$ mbar. The resulting NV centres concentration is about of $2 \times 10^{10}$ 1 cm$^2$. The depth distribution profile is nearly Gaussian, peaking at 15 nm below the surface and with a non-negligible concentration in depth ranging from ≈ 5 to 25 nm.

The plates used here are about 3 µm thick and 100 µm in size. First, a tiny drop of UV-curable glue is placed on the IC sample. The diamond with the array of plates is then brought in close proximity to the sample, with the NV layer facing toward the sample. An individual plate is then broken out of the array using a sharp tungsten tip mounted to a micromanipulator. By optical inspection, it is confirmed that the plate has not flipped during this step which means that the NV layer is facing the IC. The UV-curable glue is hardened after the plate has been fine-positioned to its final location. Finally, the Areal confocal 3D probe (NanoFocus AG) is used to measure the sample’s height profile, enabling us to estimate the thickness of the glue to 0.8(1) µm.

## Appendix B. Operational and defective devices

The micro-chip used in this article is an mm-wave test circuit for automotive radar applications designed using the multi-layered BEOL technology, developed and described by the manufacturer in [37]. A capping layer of about 2.3 µm protects the conductive layers. Adding the thickness of the glue as sample-sensor distance, we estimate the distance of the active layers to the NV centres to $\Delta z_{1,\mathrm{true}} = 4.5(1)$ µm and $\Delta z_{2,\mathrm{true}} = 7.9(1)$ µm. Although the circuit serves as a frequency doubler working in the vicinity of 160 GHz (2 × 80 GHz), it can be probed in the DC regime to evaluate the current path. To this end, expected current amplitudes for different bias voltages are given in the Supplemental Material [38]. A complete characterisation over the several bias voltages is monitored to define the operational device's functioning. Finally, comparing the experimentally extracted current amplitudes to the expected values enables us to assess the faulty behaviour.

## Appendix C. Experimental set-up and measurement

The NV imaging set-up is a custom-built wide-field fluorescence microscope similar to the one used in [24,49]. The microscope consists of an air objective (Olympus MPLAPON 50×, NA = 0.95), a 650 nm long-pass filter (Omega), a 300 nm tube lens and a Cascade II:512 CCD camera (512 × 512 pixels, Photometrics), resulting in an effective pixel width of about 192 nm on the object side. Experimental realisation of continuous-wave ODMR data was achieved by exciting NV centres with a 532 nm laser (Coherent) controlled with an acousto-optical modulator (Crystal Technology) and coupled into the optical path with a dichroic mirror (Semrock). Simultaneously, microwave radiations were generated using an MW source (SMBV100A, Rhode & Schwarz) and amplified (100S1G4, Amplifier Research) before being sent to a 50 µm-thick copper wire. The resulting MW power sent to the wire was approximately 30 dBm. For all the measurements reported in the main text, the total continuous-wave laser power at the back aperture of the objective was about 90 mW. The camera settings were set to 2 × 2 pixel-binning and the field of view (FOV) was defined to ≈ 90 µm × 90 µm. For a single frame, the exposure time was set to $t_{expo}$ = 16 ms and the frame transfer time was $t_{frame}$ = 42 ms. A single ODMR spectrum was acquired within 189 s and repeated 100 times.

The IC chip was wire bonded to a printed circuit board (PCB) with 20 µm-thick gold wires. The PCB was electrically connected to a power supply (Hameg, Rhode & Schwarz) generating 3.3 V of supply voltage to run the chip and an additional 2 V bias was used to vary the total current in the main circuit.

All measurements were performed in an ambient environment at room temperature, under a bias magnetic field $|B_0| \approx$ 5.8 mT generated using a permanent magnet thermally stabilised at a temperature of ≈ 37°C.

## Acknowledgments

We gratefully acknowledge T. Rendler for valuable discussions and technical assistance. We acknowledge the support of the European Union’s Horizon 2020 research and innovation programme under the Marie Skłodowska-Curie grant agreement N°"765267" (QuSCo) and the grant agreement N°"820394" (ASTERIQS). This work was supported by the European research council via SMeL, DFG via GRK 2642 and by the Fraunhofer Society via the project QMag.

## References

[1] Waldrop, M. M. The chips are down for Moore’s law. *Nature News* **530,** 144 (2016).

[2] Xu, Q., Schmidt, B., Pradhan, S. & Lipson, M. Micrometre-scale silicon electro-optic modulator. *Nature 2005 435:7040* **435,** 325–327 (2005).

[3] Iwai, H. & Ohmi, S. Silicon integrated circuit technology from past to future. *Microelectronics Reliability* **42,** 465–491 (2002).

[4] Riordan, M. & Hoddeson, L. Crystal Fire: The Invention, Development and Impact of the Transistor, Adapted from Chapter 1 of Crystal Fire: The Birth of the Information Age, by Michael Riordan and Lillian Hoddeson, published in 1997 by W. W. Norton & Company. *IEEE Solid-State Circuits Newsletter* **12,** 24–29 (2009).

[5] Berners-Lee, T., Cailliau, R., Luotonen, A., Nielsen, H. F. & Secret, A. The World-Wide Web. *Commun ACM* **37,** 76–82 (1994).

[6] Skog, I. & Händel, P. In-car positioning and navigation technologies - a survey. *IEEE Transactions on Intelligent Transportation Systems* **10,** 4–21 (2009).

[7] Bhatia, N. & El-Chami, M. Leadless pacemakers: a contemporary review. *J Geriatr Cardiol* **15,** 249 (2018).

[8] Moore, G. E. Cramming more components onto integrated circuits, Reprinted from Electronics, volume 38, number 8, April 19, 1965, pp.114 ff. *IEEE Solid-State Circuits Society Newsletter* **11,** 33–35 (2006).

[9] Mamaluy, D. & Gao, X. The fundamental downscaling limit of field effect transistors. *Appl Phys Lett* **106,** 193503 (2015).

[10] Bhargava, P., Kim, T., Poulton, C. V., Notaros, J., Yaacobi, A., Timurdogan, E., Baiocco, C.,

Fahrenkopf, N., Kruger, S., Ngai, T., Timalsina, Y., Watts, M. R. & Stojanovic, V. Fully Integrated Coherent LiDAR in 3D-Integrated Silicon Photonics/65nm CMOS. *IEEE Symposium on VLSI Circuits, Digest of Technical Papers* C262–C263 (2019).

[11] Chow, J., Dial, O. & Gambetta, J. IBM Quantum breaks the 100-qubit processor barrier | IBM Research Blog. *IBM Research Blog* (2021). at <https://research.ibm.com/blog/127-qubit-quantum-processor-eagle>

[12] Gargini, P., Balestra, F. & Hayashi, Y. Roadmapping of Nanoelectronics for the New Electronics Industry. *Applied Sciences* **12,** 308 (2022).

[13] Salahuddin, S., Ni, K. & Datta, S. The era of hyper-scaling in electronics. *Nat Electron* **1,** 442–450 (2018).

[14] DeBenedictis, E. P., Badaroglu, M., Chen, A., Conte, T. M. & Gargini, P. Sustaining Moore's law with 3D chips. *Computer (Long Beach Calif)* **50,** 69–73 (2017).

[15] Lancaster, A. & Keswani, M. Integrated circuit packaging review with an emphasis on 3D packaging. *Integration* **60,** 204–212 (2018).

[16] Topol, A. W., La Tulipe, D. C., Shi, L., Frank, D. J., Bernstein, K., Steen, S. E., Kumar, A., Singco, G. U., Young, A. M., Guarini, K. W. & Ieong, M. Three-dimensional integrated circuits. *IBM J Res Dev* **50,** 491–506 (2006).

[17] Wang, H., Forte, D., Tehranipoor, M. M. & Shi, Q. Probing Attacks on Integrated Circuits: Challenges and Research Opportunities. *IEEE Des Test* **34,** 63–71 (2017).

[18] Fong, L. E., Holzer, J. R., McBride, K. K., Lima, E. A., Baudenbacher, F. & Radparvar, M. High-resolution room-temperature sample scanning superconducting quantum interference device microscope configurable for geological and biomagnetic applications. *Review of Scientific Instruments* **76,** 053703 (2005).

[19] Schrag, B. D. & Xiao, G. Submicron electrical current density imaging of embedded microstructures. *Appl Phys Lett* **82,** 3272 (2003).

[20] Herrera-May, A. L., Aguilera-Cortés, L. A., García-Ramírez, P. J. & Manjarrez, E. Resonant Magnetic Field Sensors Based On MEMS Technology. *Sensors* **9,** 7785–7813 (2009).

[21] Jelezko, F. & Wrachtrup, J. Single defect centres in diamond: A review. *physica status solidi (a)* **203,** 3207–3225 (2006).

[22] Doherty, M. W., Manson, N. B., Delaney, P., Jelezko, F., Wrachtrup, J. & Hollenberg, L. C. L. The nitrogen-vacancy colour centre in diamond. *Phys Rep* **528,** 1–45 (2013).

[23] Balasubramanian, G., Chan, I. Y., Kolesov, R., Al-Hmoud, M., Tisler, J., Shin, C., Kim, C., Wojcik, A., Hemmer, P. R., Krueger, A., Hanke, T., Leitenstorfer, A., Bratschitsch, R., Jelezko, F. & Wrachtrup, J. Nanoscale imaging magnetometry with diamond spins under ambient conditions. *Nature* **455,** 648–651 (2008).

[24] Steinert, S., Dolde, F., Neumann, P., Aird, A., Naydenov, B., Balasubramanian, G., Jelezko, F. & Wrachtrup, J. High sensitivity magnetic imaging using an array of spins in diamond. *Review of Scientific Instruments* **81,** 043705 (2010).

[25] Batalov, A., Jacques, V., Kaiser, F., Siyushev, P., Neumann, P., Rogers, L. J., McMurtrie, R. L., Manson, N. B., Jelezko, F. & Wrachtrup, J. Low Temperature Studies of the Excited-State Structure of Negatively Charged Nitrogen-Vacancy Color Centers in Diamond. *Phys Rev Lett* **102,** 195506 (2009).

[26] Steinert, S., Ziem, F., Hall, L. T., Zappe, A., Schweikert, M., Götz, N., Aird, A., Balasubramanian, G., Hollenberg, L. & Wrachtrup, J. Magnetic spin imaging under ambient conditions with sub-cellular resolution. *Nat Commun* **4,** 1607 (2013).

[27] Happacher, J., Broadway, D. A., Bocquel, J., Reiser, P., Jimenéz, A., Tschudin, M. A., Thiel, L., Rohner, D., Puigibert, M. L. G., Shields, B., Maze, J. R., Jacques, V. & Maletinsky, P. Low-Temperature Photophysics of Single Nitrogen-Vacancy Centers in Diamond. *Phys Rev Lett* **128,** 177401 (2022).

[28] Liu, G. Q., Feng, X., Wang, N., Li, Q. & Liu, R. B. Coherent quantum control of nitrogen-vacancy center spins near 1000 kelvin. *Nat Commun* **10,** 1344 (2019).

[29] Lesik, M., Plisson, T., Toraille, L., Renaud, J., Occelli, F., Schmidt, M., Salord, O., Delobbe, A., Debuisschert, T., Rondin, L., Loubeyre, P. & Roch, J. F. Magnetic measurements on

micrometer-sized samples under high pressure using designed NV centers. *Science (1979)* **366,** 1359–1362 (2019).

[30] Wolf, T., Neumann, P., Nakamura, K., Sumiya, H., Ohshima, T., Isoya, J. & Wrachtrup, J. Subpicotesla diamond magnetometry. *Phys Rev X* **5,** 041001 (2015).

[31] Zhang, C., Shagieva, F., Widmann, M., Kübler, M., Vorobyov, V., Kapitanova, P., Nenasheva, E., Corkill, R., Rhrle, O., Nakamura, K., Sumiya, H., Onoda, S., Isoya, J. & Wrachtrup, J. Diamond Magnetometry and Gradiometry towards Subpicotesla dc Field Measurement. *Phys Rev Appl* **15,** 064075 (2021).

[32] Chang, K., Eichler, A., Rhensius, J., Lorenzelli, L. & Degen, C. L. Nanoscale Imaging of Current Density with a Single-Spin Magnetometer. *Nano Lett* **17,** 2367–2373 (2017).

[33] Tetienne, J. P., Dontschuk, N., Broadway, D. A., Stacey, A., Simpson, D. A. & Hollenberg, L. C. L. Quantum imaging of current flow in graphene. *Sci Adv* **3,** (2017).

[34] Nowodzinski, A., Chipaux, M., Toraille, L., Jacques, V., Roch, J. F. & Debuisschert, T. Nitrogen-Vacancy centers in diamond for current imaging at the redistributive layer level of Integrated Circuits. *Microelectronics Reliability* **55,** 1549–1553 (2015).

[35] Turner, M. J., Langellier, N., Bainbridge, R., Walters, D., Meesala, S., Babinec, T. M., Kehayias, P., Yacoby, A., Hu, E., Lončar, M., Walsworth, R. L. & Levine, E. V. Magnetic Field Fingerprinting of Integrated-Circuit Activity with a Quantum Diamond Microscope. *Phys Rev Appl* **14,** 014097 (2020).

[36] Oliver, S. M., Martynowych, D. J., Turner, M. J., Hopper, D. A., Walsworth, R. L. & Levine, E. V. Vector Magnetic Current Imaging of an 8 nm Process Node Chip and 3D Current Distributions Using the Quantum Diamond Microscope. *Conference Proceedings from the International Symposium for Testing and Failure Analysis* 96–107 (2021).

[37] Böck, J., Aufinger, K., Boguth, S., Dahl, C., Knapp, H., Liebl, W., Manger, D., Meister, T. F., Pribil, A., Wursthorn, J., Lachner, R., Heinemann, B., Rücker, H., Fox, A., Barth, R., Fischer, G., Marschmeyer, S., Schmidt, D., Trusch, A. & Wipf, C. SiGe HBT and BiCMOS process integration optimization within the DOTSEVEN project. *Proceedings of the IEEE Bipolar/BiCMOS Circuits and Technology Meeting* 121–124 (2015).

[38] Garsi, M., Stöhr, R., Denisenko, A., Shagieva, F., Trautmann, N., Vogl, U., Sene, B., Kaiser, F., Zappe, A., Reuter, R. & Wrachtrup, J. Three-dimensional integrated circuit activity imaging using quantum defects in diamond. See Supplemental Material, p. S1- p. S26, https://journals.aps.org/prapplied/supplemental/10.1103/PhysRevApplied.21.014055

[39] Roth, B. J., Sepulveda, N. G. & Wikswo, J. P. Using a magnetometer to image a two-dimensional current distribution. *J Appl Phys* **65,** 361–372 (1989).

[40] Meltzer, A. Y., Levin, E. & Zeldov, E. Direct Reconstruction of Two-Dimensional Currents in Thin Films from Magnetic-Field Measurements. *Phys Rev Appl* **8,** 064030 (2017).

[41] Burkacky, O., Patel, M., Sergeant, N. & Thomas, C. Reimagining fabs: Advanced analytics in semiconductor manufacturing. *Copyright © 2017 McKinsey & Company.* 7

[42] Pierce, D. G. & Brusius, P. G. Electromigration: A review. *Microelectronics Reliability* **37,** 1053–1072 (1997).

[43] Singh, N., Bower, A. F. & Shankar, S. A three-dimensional model of electromigration and stress induced void nucleation in interconnect structures. *Model Simul Mat Sci Eng* **18,** 065006 (2010).

[44] Ceric, H. & Selberherr, S. Electromigration in submicron interconnect features of integrated circuits. *Materials Science and Engineering: R: Reports* **71,** 53–86 (2011).

[45] Bishop, C. M. *Pattern Recognition and Machine Learning*. (Springer New York, 2006).

[46] Holler, M., Odstrcil, M., Guizar-Sicairos, M., Lebugle, M., Müller, E., Finizio, S., Tinti, G., David, C., Zusman, J., Unglaub, W., Bunk, O., Raabe, J., Levi, A. F. J. & Aeppli, G. Three-dimensional imaging of integrated circuits with macro- to nanoscale zoom. *Nat Electron* **2,** 464–470 (2019).

[47] Chipaux, M., Toraille, L., Larat, C., Morvan, L., Pezzagna, S., Meijer, J. & Debuisschert, T. Wide bandwidth instantaneous radio frequency spectrum analyzer based on nitrogen vacancy centers in diamond. *Appl Phys Lett* **107,** 233502 (2015).

[48] Zopes, J. & Degen, C. L. Reconstruction-Free Quantum Sensing of Arbitrary Waveforms. *Phys Rev Appl* **12,** 054028 (2019).
[49] Ziem, F., Garsi, M., Fedder, H. & Wrachtrup, J. Quantitative nanoscale MRI with a wide field of view. *Sci Rep* **9,** 12166 (2019).
[50] Mizuno, K., Ishiwata, H., Masuyama, Y., Iwasaki, T. & Hatano, M. Simultaneous wide-field imaging of phase and magnitude of AC magnetic signal using diamond quantum magnetometry. *Sci Rep* **10,** 11611 (2020).
[51] Casola, F., Van Der Sar, T. & Yacoby, A. Probing condensed matter physics with magnetometry based on nitrogen-vacancy centres in diamond. *Nat Rev Mater* **3,** 17088 (2018).
[52] Webb, J. L., Troise, L., Hansen, N. W., Frellsen, L. F., Osterkamp, C., Jelezko, F., Jankuhn, S., Meijer, J., Berg-Sørensen, K., Perrier, J. F., Huck, A. & Andersen, U. L. High-Speed Wide-Field Imaging of Microcircuitry Using Nitrogen Vacancies in Diamond. *Phys Rev Appl* **17,** 064051 (2022).
[53] Parashar, M., Bathla, A., Shishir, D., Gokhale, A., Bandyopadhyay, S. & Saha, K. Sub-second Temporal Magnetic Field Microscopy Using Quantum Defects in Diamond. *Sci Rep* **12,** 8743 (2022).
[54] Abrahams, G. J., Scholten, S. C., Healey, A. J., Robertson, I. O., Dontschuk, N., Lim, S. Q., Johnson, B. C., Simpson, D. A., Hollenberg, L. C. L. & Tetienne, J. P. An integrated widefield probe for practical diamond nitrogen-vacancy microscopy. *Appl Phys Lett* **119,** 254002 (2021).
[55] Toraille, L., Hilberer, A., Plisson, T., Lesik, M., Chipaux, M., Vindolet, B., Pépin, C., Occelli, F., Schmidt, M., Debuisschert, T., Guignot, N., Itié, J. P., Loubeyre, P. & Roch, J. F. Combined synchrotron x-ray diffraction and NV diamond magnetic microscopy measurements at high pressure. *New J Phys* **22,** 103063 (2020).
[56] Courjault, N., Infante, F., Bley, V., Lebey, T. & Perdu, P. Improvement of 3D current mapping by coupling magnetic microscopy and X-Ray computed tomography. in *Proceedings of the 21th International Symposium on the Physical and Failure Analysis of Integrated Circuits (IPFA)* 26–29 (2014).
[57] Eroğlu, H. H., Puonti, O., Göksu, C., Gregersen, F., Siebner, H. R., Hanson, L. G. & Thielscher, A. On the reconstruction of magnetic resonance current density images of the human brain: Pitfalls and perspectives. *Neuroimage* **243,** 118517 (2021).
[58] Yang, B. J., Moon, E. G., Isobe, H. & Nagaosa, N. Quantum criticality of topological phase transitions in three-dimensional interacting electronic systems. *Nature Physics 2014 10:10* **10,** 774–778 (2014).
[59] Wang, C., Potter, A. C. & Senthil, T. Classification of interacting electronic topological insulators in three dimensions. *Science (1979)* **343,** 629–630 (2014).

## Figures

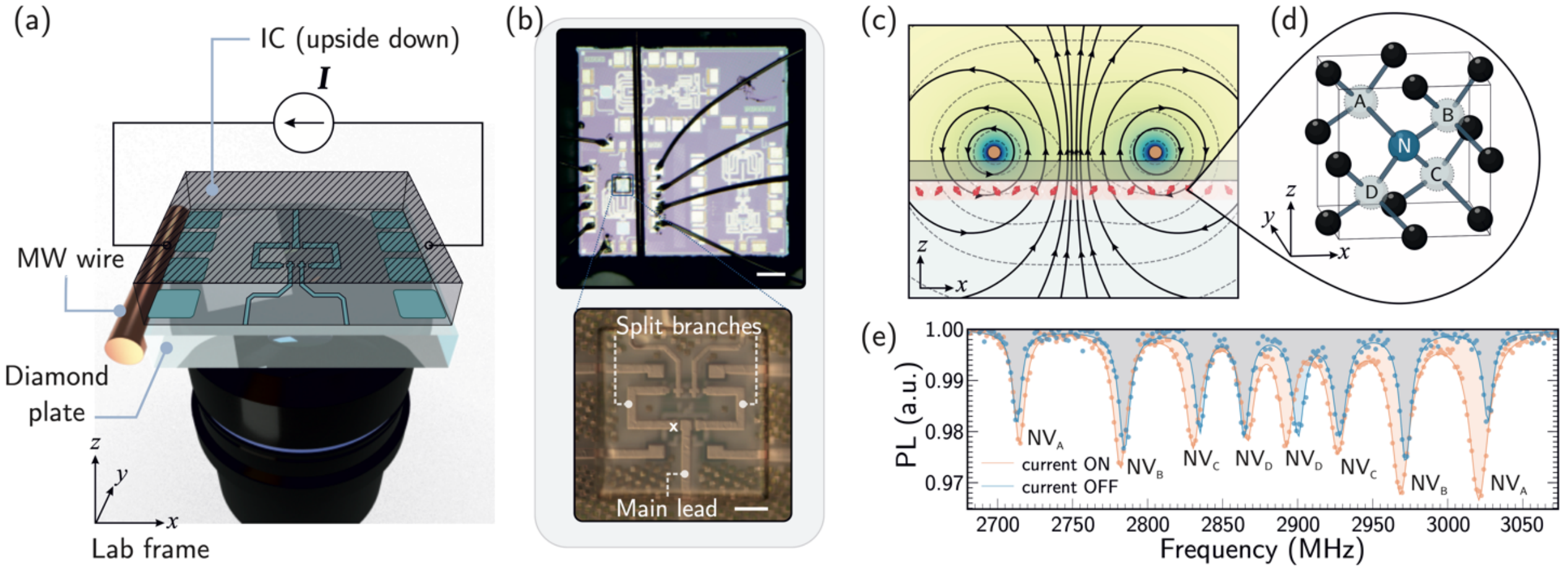

**Fig. 1. Mapping integrated circuit activity with quantum sensors. a)** Schematic of the experiment. A microfabricated diamond plate contains a layer of near-surface NV centres and is glued to an integrated circuit. The sample is mounted to an inverted microscope where laser and microwave radiations (MWs) excite the NV centres. A CCD camera records the emitted photoluminescence (PL). **b)** Photograph of the IC microchip. The upper picture shows the overall chip with different circuit designs. The diamond plate is glued to a region of interest outlined by a blue square. A copper wire carrying MWs is placed next to the diamond, and wire bonds connect the chip to a power supply. The bottom picture shows a zoom-in on the diamond plate. Scale bars are 200 µm for the upper photograph and 20 µm for the bottom one. **c)** Visualisation of the cross-section of the experiment. The current-carrying wires generate Oersted fields sensed by a layer of NV centres represented in red and separated by a protective overcoat from the leads. Solid lines with arrows represent the magnetic flux lines, and dashed lines represent magnetic field isolines. **d)** Representation of the four possible tetrahedral orientations of the NV bond (A, B, C, D) in the reference frame *xyz*. **e)** ODMR spectra from a single-pixel near the edge of a semiconductor stripe indicated by the white cross in (b). The blue spectrum is obtained with a bias magnetic field $B_0$ used to split the eight resonances lines of the NV ensemble. The orange spectrum is acquired when current flows in the IC, creating a shift in the resonances due to the Zeeman interaction of the NV centres with the Oersted field. Solid lines are multiple-Lorentzian fits. Each resonance is labelled according to the corresponding NV orientation, defined in (d).

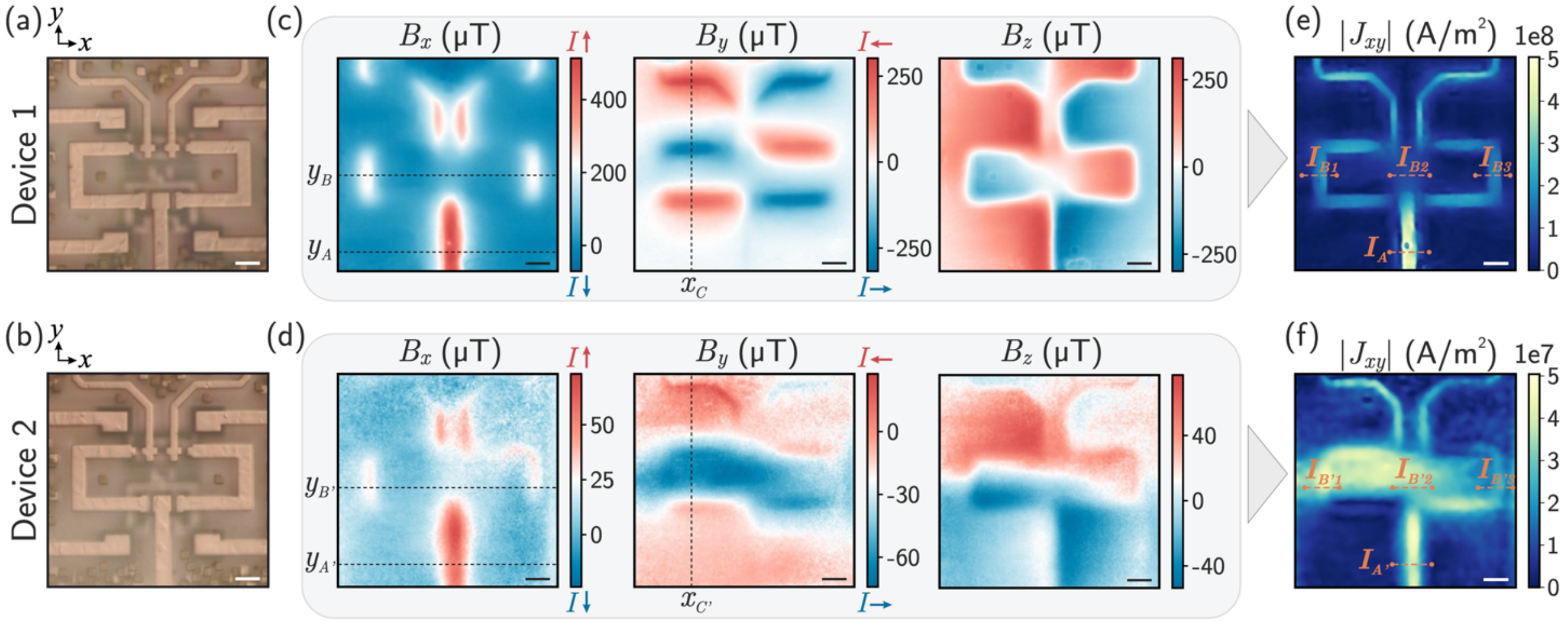

**Fig. 2. Vectorial Magnetic field produced by the current-carrying wires: device 1 vs device 2 and corresponding current density maps. a-b)** Optical images of the device 1 and device 2, respectively. **c-d)** Mapping of the three vectorial magnetic field components $B_x$, $B_y$, $B_z$ produced by the operational and defective IC, respectively. The sign gives the direction of the field. Linecuts at $y_A$, $y_{A'}$, $y_B$, $y_{B'}$ are shown in Fig. 3, b and c for further analysis. Linecuts at $x_C$, $x_{C'}$ are shown in the Supplemental Material [38]. **e-f)** Corresponding in-plane current density map reconstructed from $B_x$ and $B_y$ in (a) and (b) respectively. The orange labels indicate the sections where the current amplitudes in Fig. 3 are estimated. Scale bars are 10 µm wide.

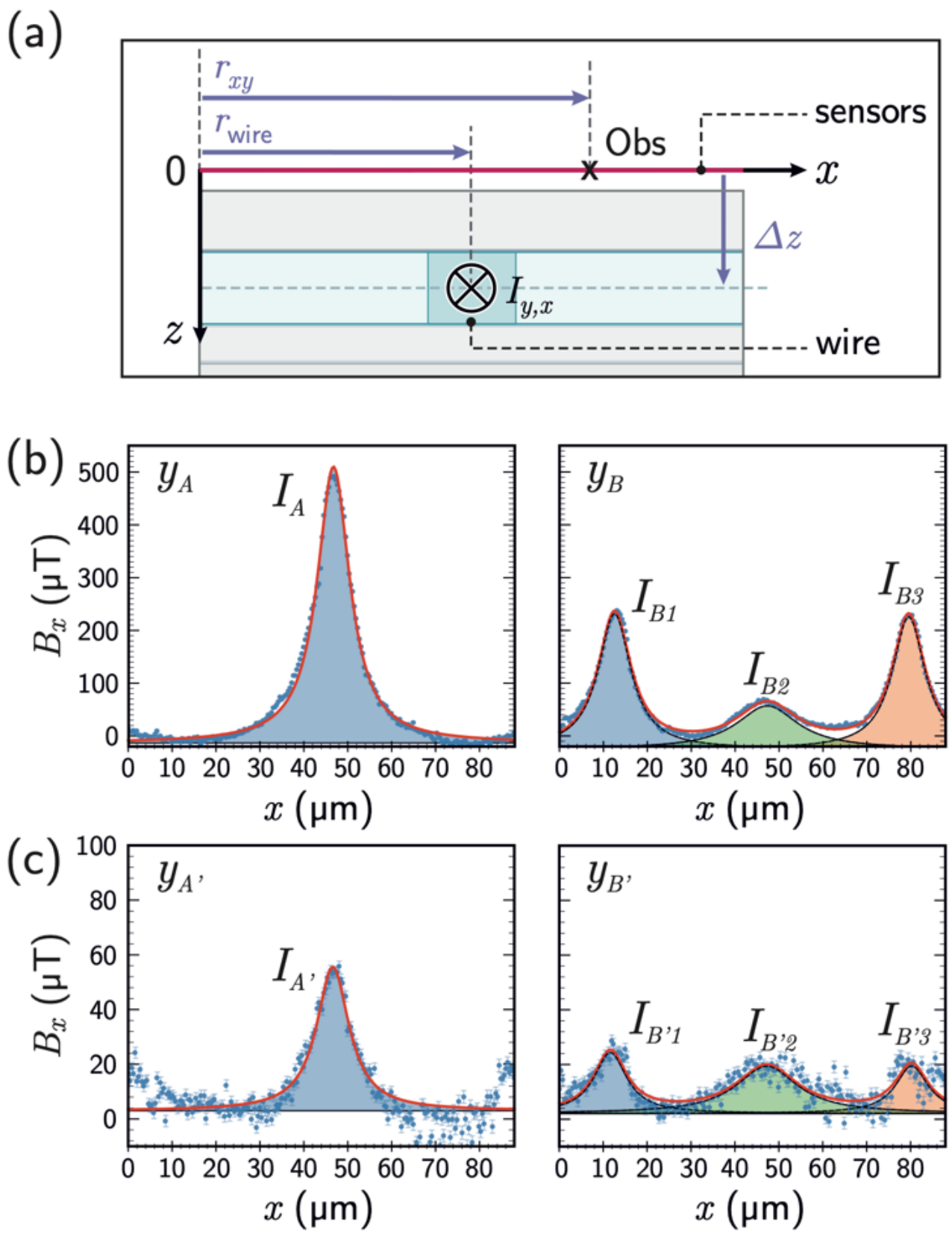

**Fig. 3. Experimental contributions of Oersted fields originating from different layers. a)** Schematic view of the *xz*-plane of the experiment. The variables used to fit the data with Eq. (2) are labelled according to an arbitrarily chosen observation point Obs. **b)** Linecuts of experimental data (blue dots) outlined in Fig. 2c along the *x*-axis at a vertical position $y_A$ (left panel) and at a vertical position $y_B$ (right panel). The fit (solid red line) returns a current amplitude of $I_A$ = 11.77(6) mA where the source-sensor distance is fixed to $\Delta z_A$ = 4.5 µm for the left panel and current amplitudes of $I_{B1}$ = 5.64(5) mA, $I_{B2}$ = 3.49(8) mA, $I_{B3}$ = 5.76(5) mA where the source-sensor distances are fixed to $\Delta z_{B1,B3}$ = 4.5 µm and $\Delta z_{B2}$ = 8.5 µm for the right panel. c) Linecuts of experimental data (blue dots) outlined in Fig. 2d along the *x*-axis at a vertical position $y_{A'}$ (left panel) and at a vertical position $y_{B'}$ (right panel). The fit (solid red line) returns a current amplitude $I_{A'}$ = 1.18(4) mA where the source-sensor distance is fixed to $\Delta z_{A'}$ = 4.5 µm for the left panel and current amplitudes of $I_{B'1}$ = 0.50(3) mA, $I_{B'2}$ = 0.73(5) mA, $I_{B'3}$ = 0.39(3) mA where the source-sensor distances are fixed to $\Delta z_{B'1,B'3}$ = 4.5 µm and $\Delta z_{B'2}$ = 8.5 µm for the right panel. Error bars correspond to one standard deviation. Plain colours underline the contribution of each single wire.

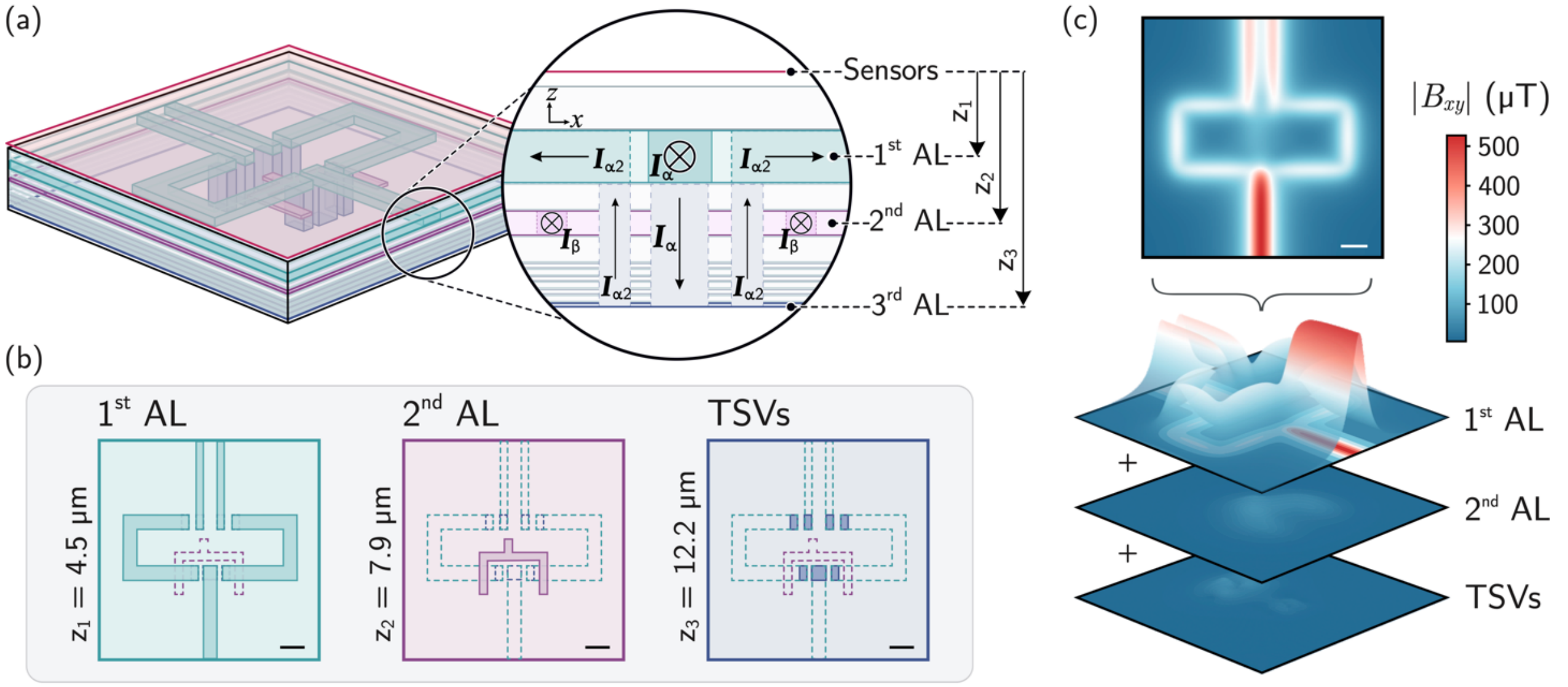

**Fig. 4. Simulation of Oersted field contributions originating from different layers. a)** Geometry of the simulated structure. A layer of NV centres is separated from the chip by 0.8 µm. The structure is composed of twelve layers comprising the two AL and TSVs. A current of amplitude $I_\alpha$ = 11.8 mA goes to the main branch of the 1st AL, flows down to the bottom layer of the structure where it splits into two sub-path with an amplitude of $I_{\alpha 2} = I_\alpha/2$ and flows back to the 1st AL. In the second AL, a current of amplitude $I_\beta$ = 2 mA is injected into each of the two branches which combine to a single one afterwards. **b)** Top view of each AL of the structure. The 1st, 2nd AL and the bottom layer of the structure are located at $z_1$ = 4.5 µm, $z_2$ = 7.9 µm and $z_3$ = 12.2 µm respectively from the sensors. **c)** *Top*: Oersted field in the *xy*-plane generated by all active components at the sensors layer position. *Bottom*: Separate contribution from each AL where the vertical axis shows the lateral magnetic field amplitude $|B_{xy}|$.

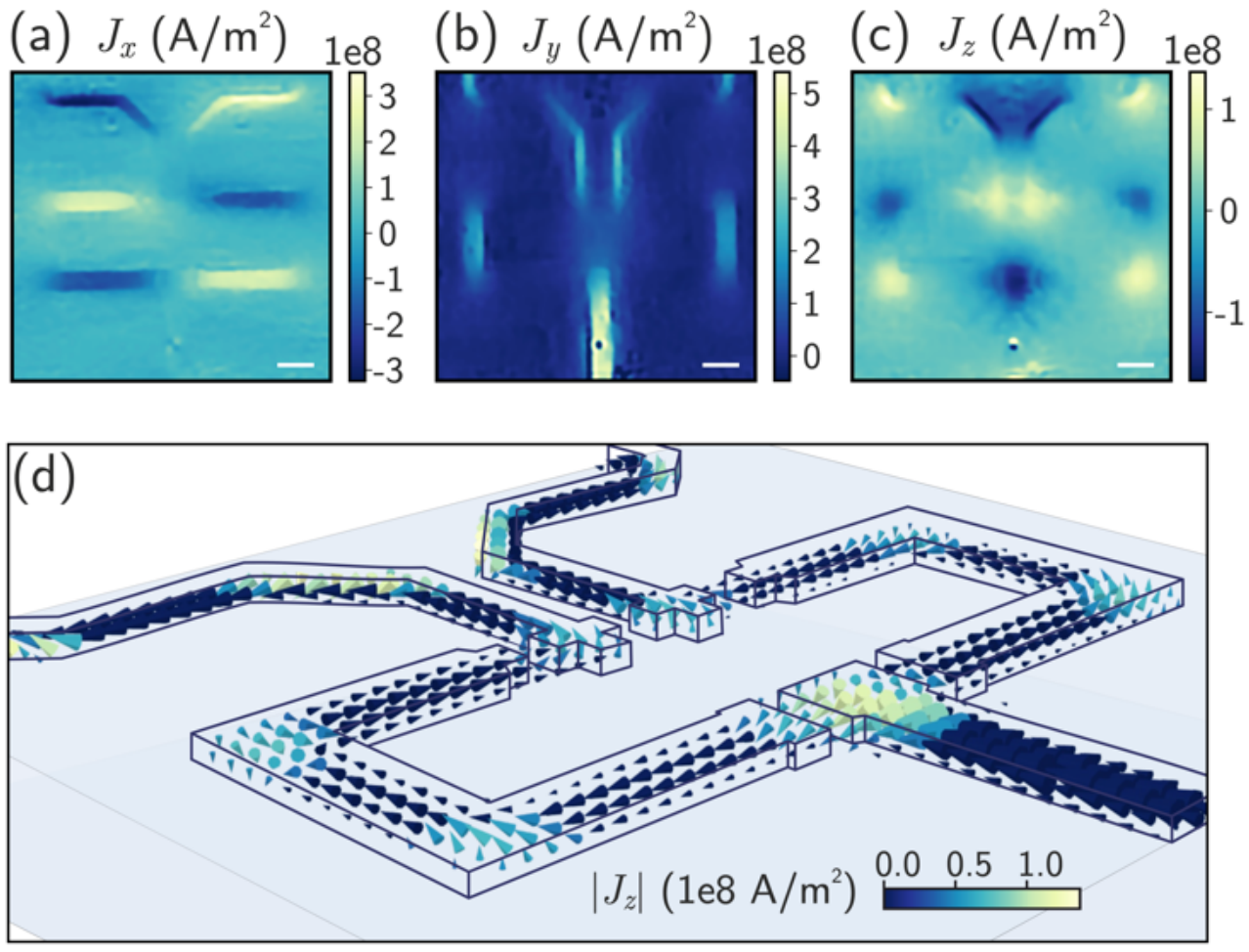

**Fig. 5. Current density maps. a-c)** Images of the three vectorial components of the current density $J_x$, $J_y$, $J_z$ for device 1. Scale bars are 10 µm wide. **d)** Three-dimensional representation of the current flow in the outer layer of the IC. The thickness of the arrow scales with the total current density magnitude and the colour scales with the magnitude of $J_z$.

Supplemental Material for

# Three-dimensional imaging of integrated-circuit activity using quantum defects in diamond

Marwa Garsi[1,2]*, Rainer Stöhr[1], Andrej Denisenko[2], Farida Shagieva[2], Nils Trautmann[3], Ulrich Vogl[3], Badou Sene[4], Florian Kaiser[1†], Andrea Zappe[1], Rolf Reuter[1], Jörg Wrachtrup[1]

1) 3rd Institute of Physics, IQST, and Research Center SCoPE, University of Stuttgart, 70569 Stuttgart, Germany.
2) SQUTEC-TTI GmbH, 70569 Stuttgart, Germany.
3) Corporate Research and Technology, Carl Zeiss AG, Carl-Zeiss-Strasse 22, 73447 Oberkochen, Germany.
4) Mobility Electronics, Robert Bosch GmbH, 72762 Reutlingen, Germany.

† Now at Luxembourg Institute of Science and Technology (LIST)

* m.garsi@pi3.uni-stuttgart.de

## Contents

## S1. Measurement of the electron ground state Hamiltonian of the NV centre

### S1.1. ODMR acquisition of an ensemble of NV centres

Probing the electron ground-state Hamiltonian of an ensemble of NV centres provides the complete vector information about the applied magnetic field through the Zeeman interaction term. The measurement consists of recording the photoluminescence (PL) intensity of the NV centres as a function of an applied MW frequency to form an optically detected magnetic resonance (ODMR) spectrum (Fig. S1). We use a bias magnetic field $B_0$ to lift the eight otherwise degenerate electron spin resonances and identify the four crystallographic NV orientations. The amplitude of this bias field satisfies $|B_I| << |B_0|$, where $B_I$ is the current-induced field. Hence, $B_I$ manifests small shifts in the ODMR frequencies, as illustrated in Fig. 1e of the main text. To account for the Oersted field created by the closed-circuit only, we map the difference between the magnetic field obtained with closed-circuit and the magnetic field obtained with open circuit.

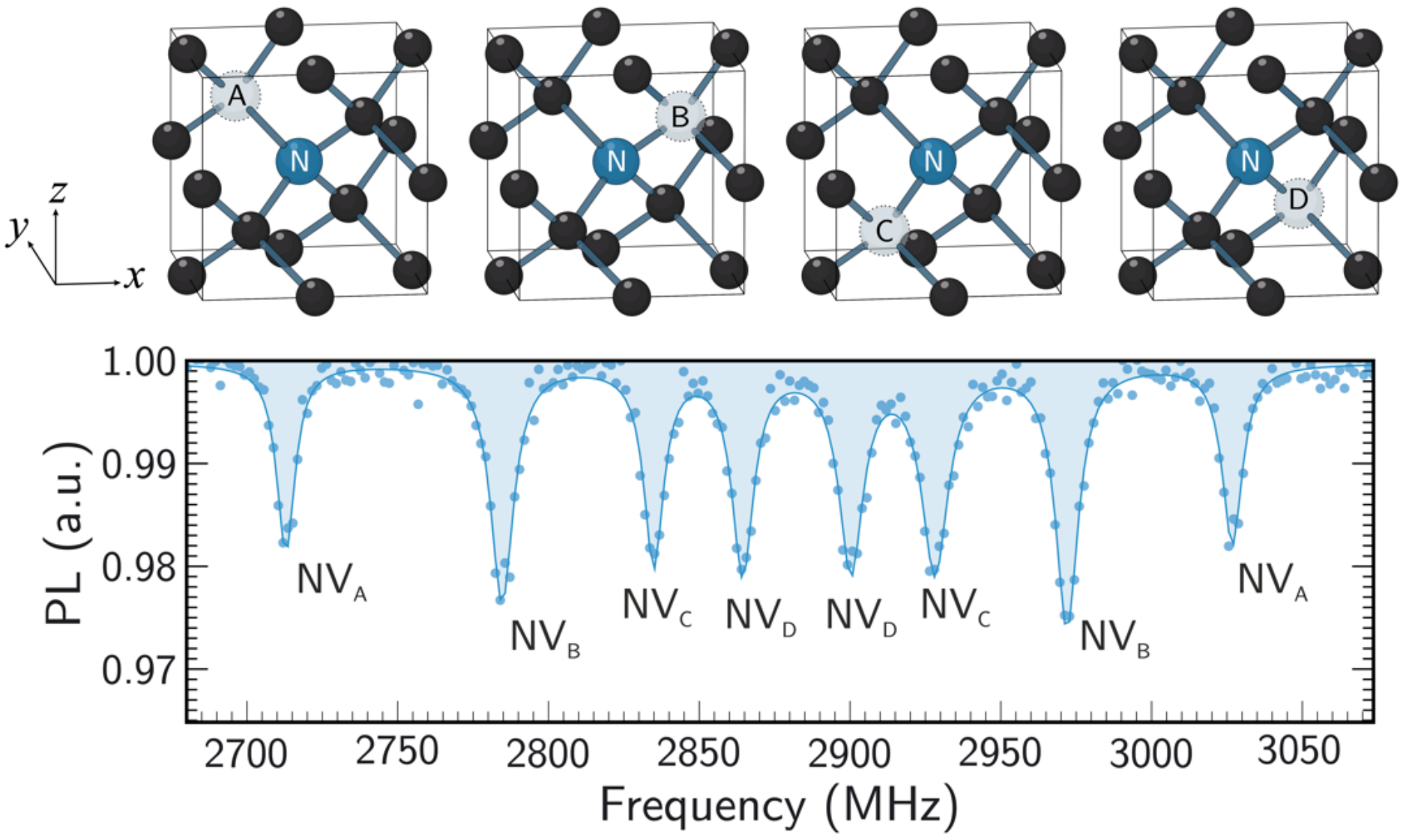

**Fig. S1. ODMR spectrum of an ensemble of NV centres.** The ODMR spectrum shows the Zeeman splitting of the four crystallographic NV orientations with an arbitrarily chosen static magnetic field $B_0$. The four possible NV-associated axes are defined by the combination of the Nitrogen atom labelled with the letter N with one of the vacancies labelled with the letters A, B, C, D.

The integrated circuit's chip is made of SiGe [37] and can emit fluorescence in our detection window [60]. To suppress the chip's illumination effects and fluctuations of the ODMR baseline, we acquire the ODMR signal with microwave ON labelled as $PL_{ON}$ and with microwave OFF labelled as $PL_{OFF}$. We use $PL_{ON}$ and $PL_{OFF}$ to compute the Michelson Contrast of the ODMR signal, defined as:

$$\text{Michelson Contrast} = \frac{PL_{ON} - PL_{OFF}}{PL_{ON} + PL_{OFF}} \tag{S1}$$

## S1.2. Fitting procedure

In order to extract the full vector information from the ODMR spectrum for each pixel, we adopted a similar strategy than in [24]. We developed an optimised procedure based on quasi-Newton methods to ensure a reliable determination of several parameters [61]. The procedure, as shown in Fig. S2 is as follows:

First, we use a nonlinear least-squares numerical minimisation method to fit the ODMR spectrum with eight Lorentzian functions, defined as:

$$L(\omega; A, \omega^{\text{exp}}, \sigma) = y_0 + \sum_{i=A}^{D} \sum_{j=\pm} \frac{A_{i,j}}{\pi} \left[ \frac{\sigma_{i,j}}{\left(\omega - \omega_{i,j}^{\text{exp}}\right)^2 + \sigma_{i,j}^2} \right] \tag{S2}$$

where the resonance frequencies $\omega_{i,j}^{\text{exp}}$, linewidths $2\sigma_{i,j}$, amplitudes $A_{i,j}$ and the offset $y_0$ are all free parameters. The indices *{i,j}* identify the resonance frequencies where $\mathrm{i} \in [\mathrm{A}, \mathrm{B}, \mathrm{C}, \mathrm{D}]$ is associated to the NV orientation as illustrated in Fig. S1 and $\mathrm{j} \in [-, +]$ is associated to the spin states $\mathrm{m}_s = -1$ and $m_s = +1$ respectively.

For a robust analysis, we use a fitting procedure based on a Levenberg-Marquardt algorithm [62,63]. The Levenberg-Marquardt algorithm reduces the sum of the squares of the errors between the model function defined by Eq.(S2) and the data points acquired experimentally through a sequence of well-chosen values to update the model parameters. With such an algorithm, the choice of the initial guess is crucial to allow the minimisation to converge towards the right solution. As the resonance shifts are small across the field of view (< 1%), we chose to compute the median of the ODMR spectra over all the pixels and take the returned fitting parameters as the initial guess for each pixel.

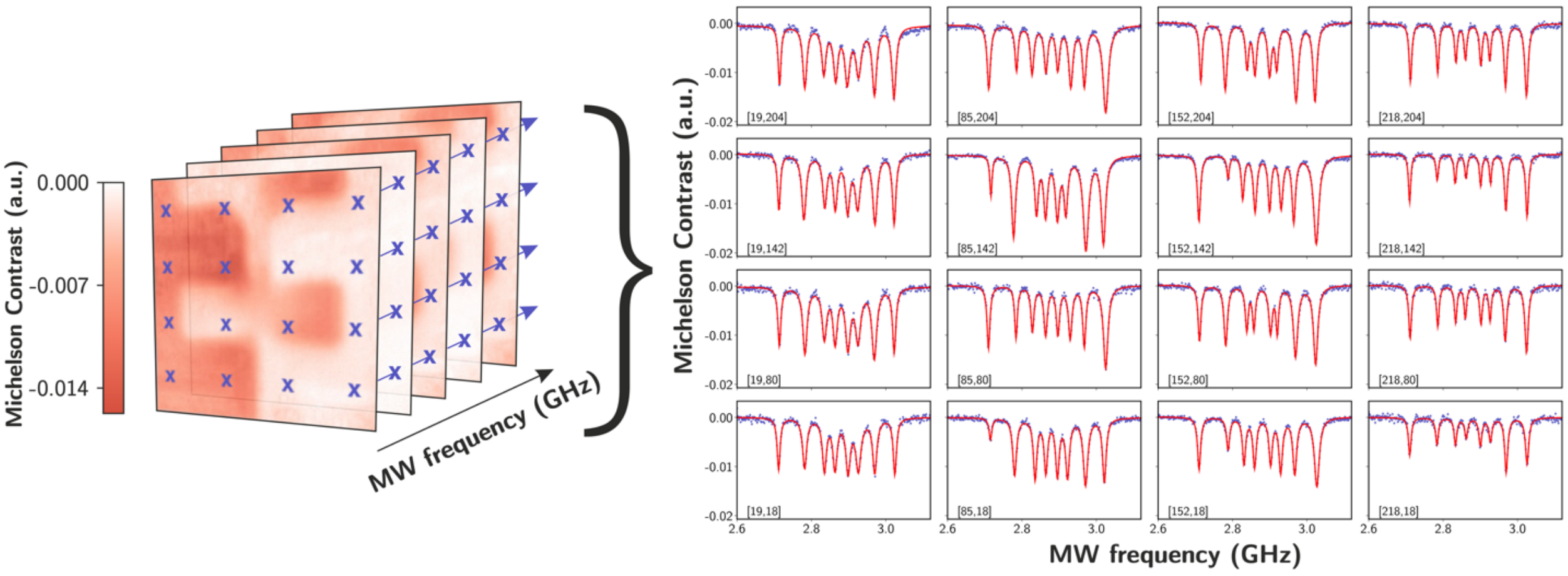

**Fig. S2. ODMR fitting procedure.** Each pixel contains an ODMR spectrum fitted using a Levenberg-Marquardt algorithm. *Left:* Michelson contrast frames acquired while swapping MW frequencies. *Right:* Corresponding ODMR spectra, taken at the blue crosses positions.

From the returned fit parameters of each pixel, we extract the centre frequencies of each experimental resonance transition $\{\omega_{i,\pm}^{\text{exp}}\}_{i=A \text{ to } D}$. The resulting resonance transitions are shown in Fig. S3 for the running device and in Fig. S4 for the background measurement. The parameters $\{\omega_{i,\pm}^{\text{exp}}\}_{i=A \text{ to } D}$ are sufficient to reconstruct the vectorial magnetic field as shown in Section S1.3. However, it is worth noting that the remaining parameters can help to seek further information about the device and experimental conditions. In particular, the linewidth of the Lorentzian function depends on the MW power as derived in [32,64]. Using this information, we map the half-width at half maximum (hwhm) parameters for the background measurement to seek MW effects on the device, such as potential cross-talks. The resulting maps are shown in Fig. S5 where we can see a gradient due to the MW wire's geometry but no relevant cross-talks since the MW power is kept minimal.

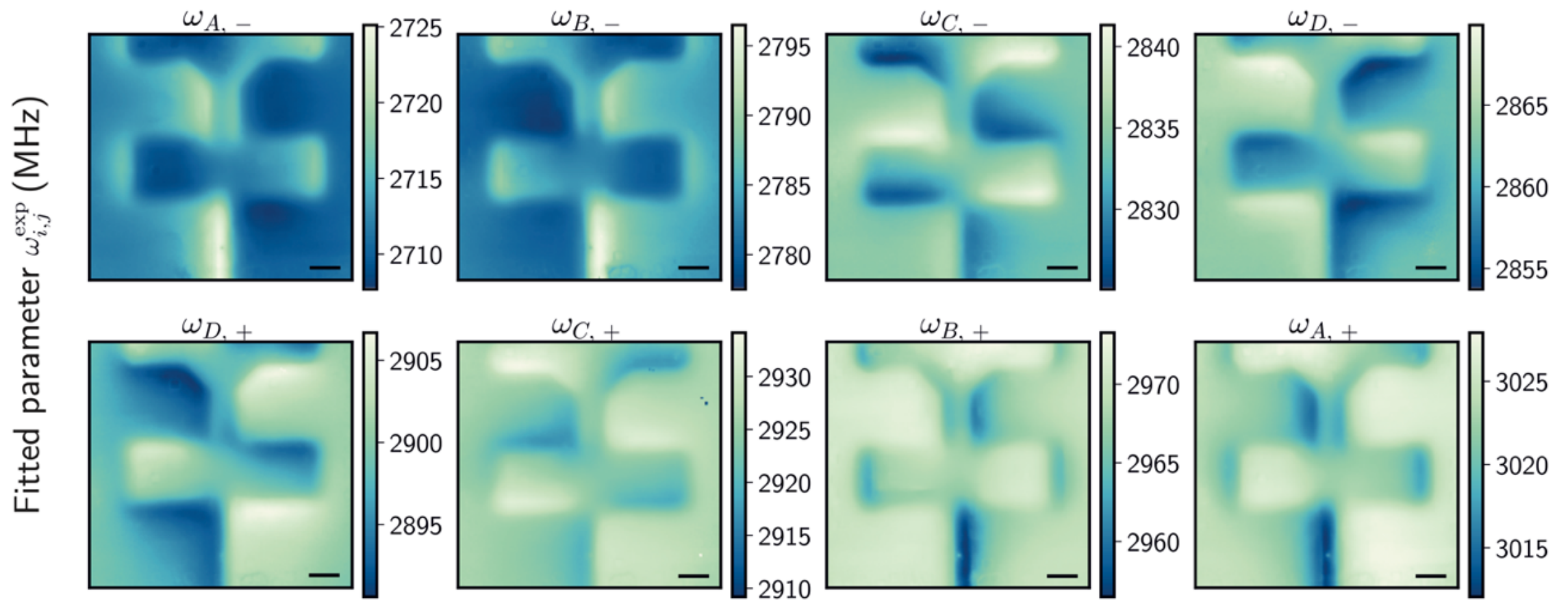

**Fig. S3. Fitted centre frequencies $\omega_{i,j}^{\text{exp}}$ for the running device's measurement.** Scale bars are 10 µm wide.

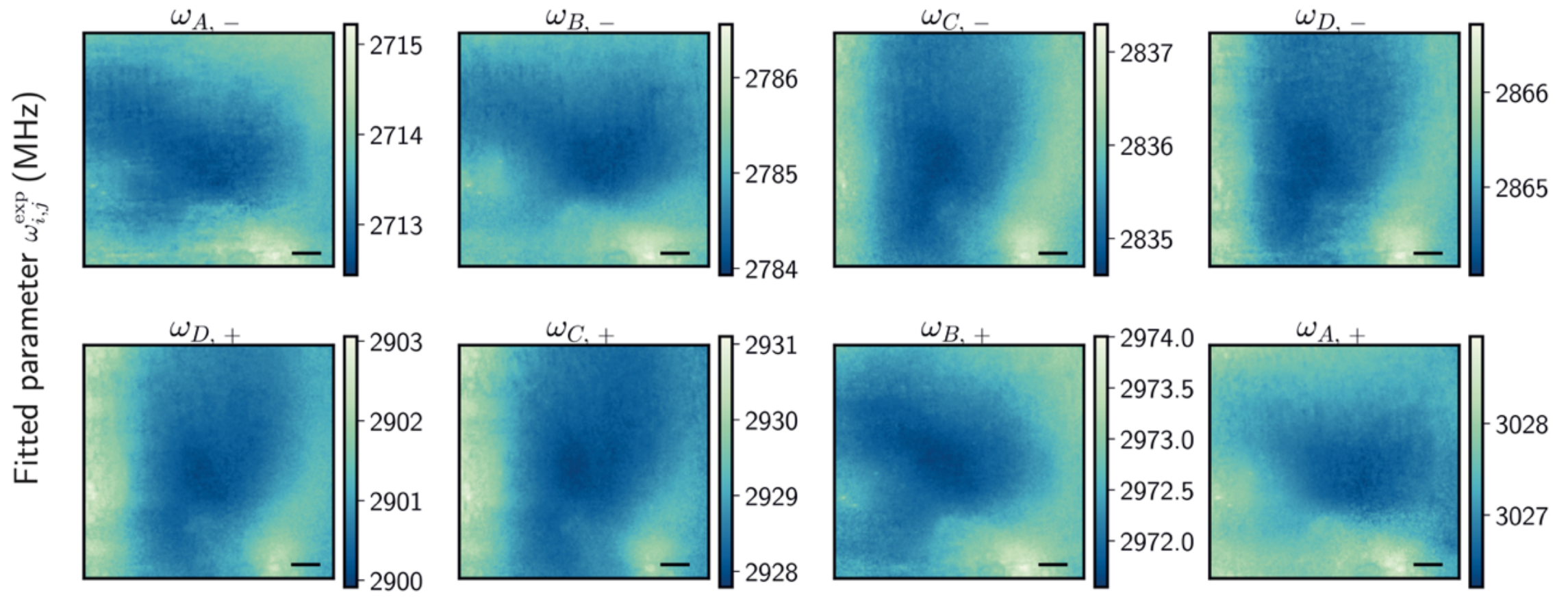

**Fig. S4. Fitted centre frequencies $\omega_{i,j}^{\text{exp}}$ for the background's measurement.** Scale bars are 10 µm wide.

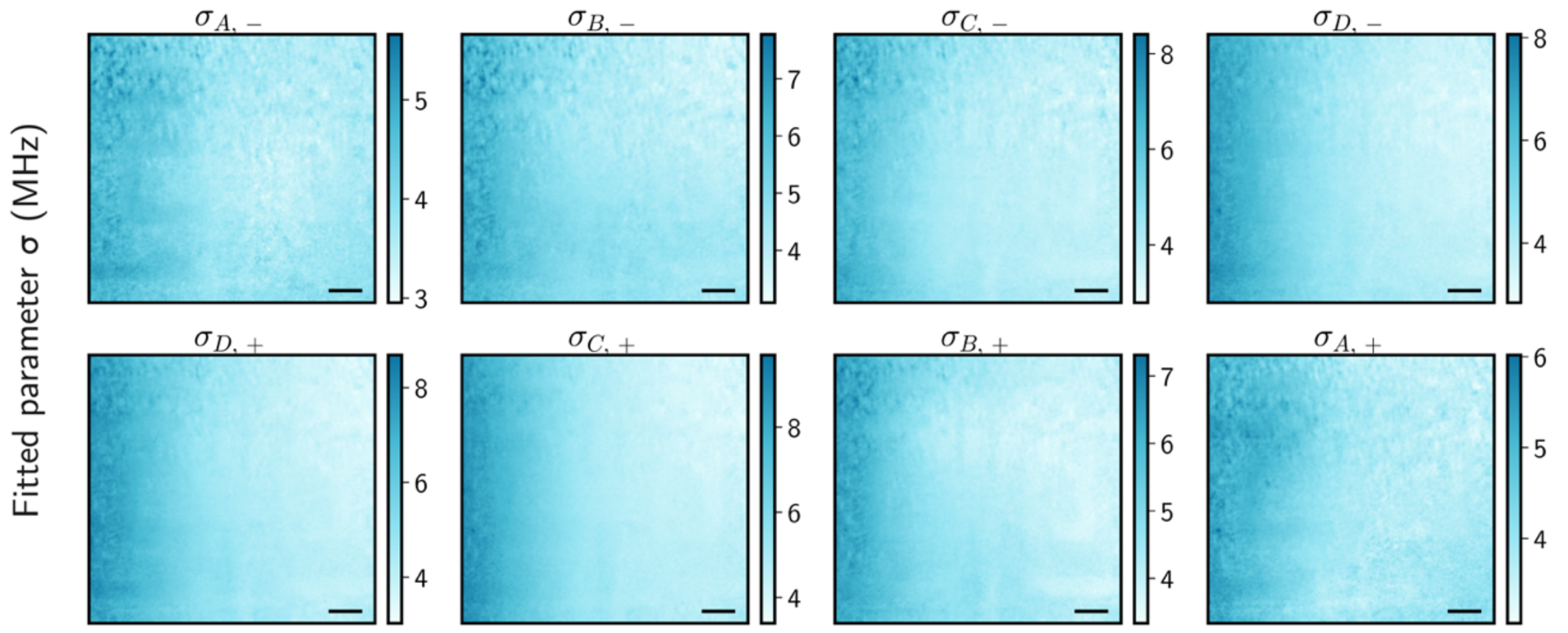

**Fig. S5. Fitted hwhm $\sigma_{i,j}$ for the background measurement.** Scale bars are 10 µm wide.

### S1.3. Ground state Hamiltonian evaluation of the NV centres electronic spin

Once the experimental resonance frequencies are known, we calculate the theoretically expected resonances $\{\omega_{i,\pm}^{\text{th}}\}_{i=A\text{ to }D}$. To this end, we solve the eigenvalues of the diagonalised ground state Hamiltonian $\mathcal{H}_{NVi}$ (Eq. {(S3), (S4), (S5)}) for each of the NV axis $\hat{u}_{NVi}$ defined by the Eq. (S6), (S7), (S8) and (S9).

NV centres are sensitive to strain, electric fields, and temperature. The Hamiltonian for the NV ground state spin $\mathcal{H}_{NVi}$ in a (100) crystal in the lab-frame *xyz*, with the presence of a bias magnetic field, an electric field and an intrinsic crystal strain is given by [22,65] :

$$\frac{\mathcal{H}_{NVi}}{\hbar} = (D_{\text{gs}} + \delta)\left[(\hat{\mathrm{u}}_{\mathrm{NV}i} \cdot \hat{\mathrm{S}}_z)^2 - \frac{2}{3}\right] + \gamma_{\mathrm{NV}}\, \vec{B} \cdot \vec{S} \tag{S3}$$

$$- \alpha\left[(\hat{\mathrm{u}}_{\mathrm{NV}i} \cdot \hat{\mathrm{S}}_x)^2 - (\hat{\mathrm{u}}_{\mathrm{NV}i} \cdot \hat{\mathrm{S}}_y)^2\right] \tag{S4}$$

$$+ \beta\left[(\hat{u}_{NVi} \cdot \hat{S}_x)(\hat{u}_{NVi} \cdot \hat{S}_y) + (\hat{u}_{NVi} \cdot \hat{S}_y)(\hat{u}_{NVi} \cdot \hat{S}_x)\right] \tag{S5}$$

where $D_{\text{gs}}$ is the temperature-dependant [66,67] zero-field splitting (ZFS) between the ground state spin sublevels, $\gamma_{\mathrm{NV}} = 28.025$ MHz/mT is the electron spin gyromagnetic ratio, $\vec{B}$ is the total magnetic field in the vicinity of the NV centre and $\vec{\mathrm{S}} = (\mathrm{S_x} \quad \mathrm{S_y} \quad \mathrm{S_z})$ represents the dimensionless electronic spin-1 operator. The parameters $\delta, \alpha, \beta$ account for the Stark shifts and can translate strain and electric effects.

In the lab frame *xyz*, the four NV axes in a (100) crystal are defined as follow:

$$\hat{u}_{NVA} = \pm\frac{1}{\sqrt{3}}\begin{pmatrix}-\sqrt{2} & 0 & 1\end{pmatrix} \quad \text{(S6)}$$

$$\hat{u}_{NVB} = \pm\frac{1}{\sqrt{3}}\begin{pmatrix}\sqrt{2} & 0 & 1\end{pmatrix} \quad \text{(S7)}$$

$$\hat{u}_{NVC} = \pm\frac{1}{\sqrt{3}}\begin{pmatrix}0 & -\sqrt{2} & 1\end{pmatrix} \quad \text{(S8)}$$

$$\hat{u}_{NVD} = \pm\frac{1}{\sqrt{3}}\begin{pmatrix}0 & \sqrt{2} & 1\end{pmatrix} \quad \text{(S9)}$$

Finally, as summarised in Fig. S6, we minimise the function:

$$\chi = \sum\left[\omega_{i,j}^{\text{exp}} - \omega(\vec{B}, D_{\text{gs}} + \delta, \alpha, \beta)\right]^2 \quad \text{(S10)}$$

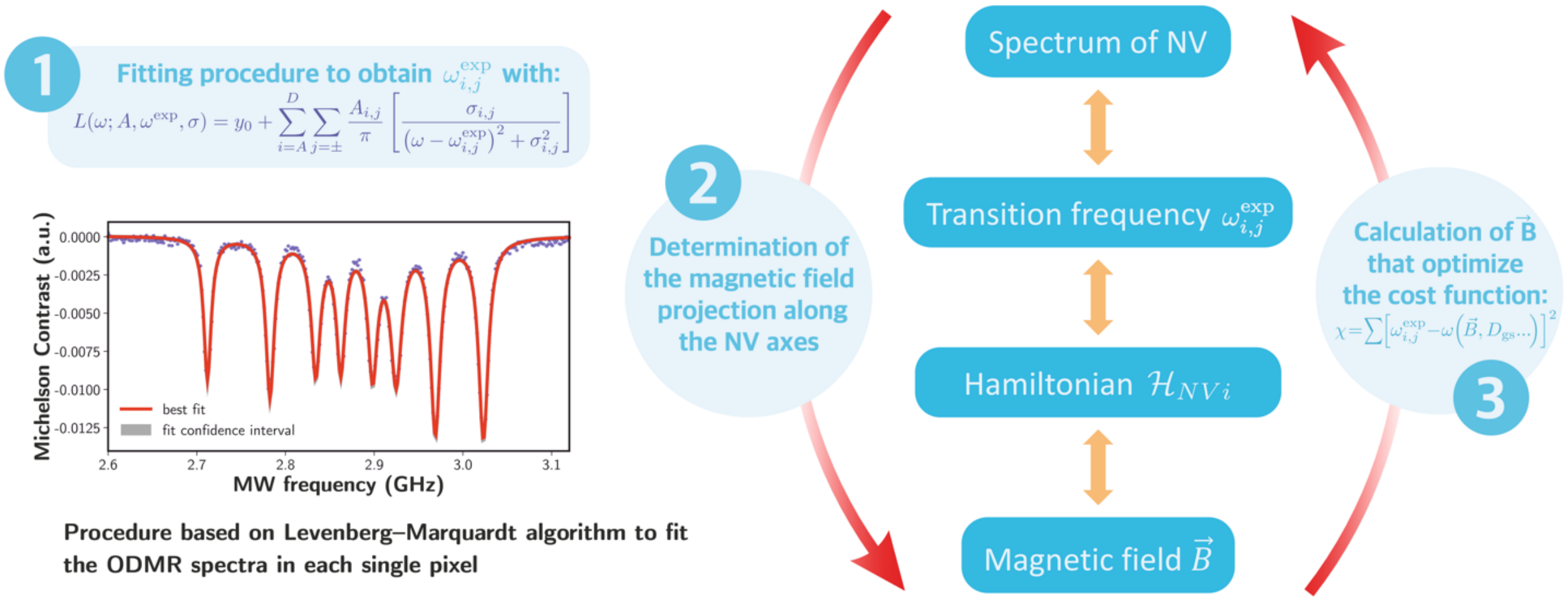

**Fig. S6. Overall procedure to extract magnetic field information.** 1) ODMR spectra are fitted for each pixel over the FOV, and the experimental resonance frequencies are extracted. 2-3) The parameters, including $\vec{B}$, are estimated by iteration until minimising the cost function χ.

In our experiment, the dominant terms in $\mathcal{H}_{NVi}$ (Eq. {(S3), (S4), (S5)}) are the ZFS and the Zeeman interaction. The resulting magnetic fields extracted from the Zeeman interaction in both devices are shown in Fig. 2 of the main text.

The ZFS has a dependence with the temperature $T$ such as $\partial D_{\text{gs}}/\partial T \approx -75$ kHz/K at room temperature [66]. With no further information about the strain inside the crystal or prior information about the temperature distribution, the terms $D_{\text{gs}}$ and $\delta$ cannot be decorrelated. The resulting map $D_{\text{gs}} + \delta$ are shown in Fig. S7.

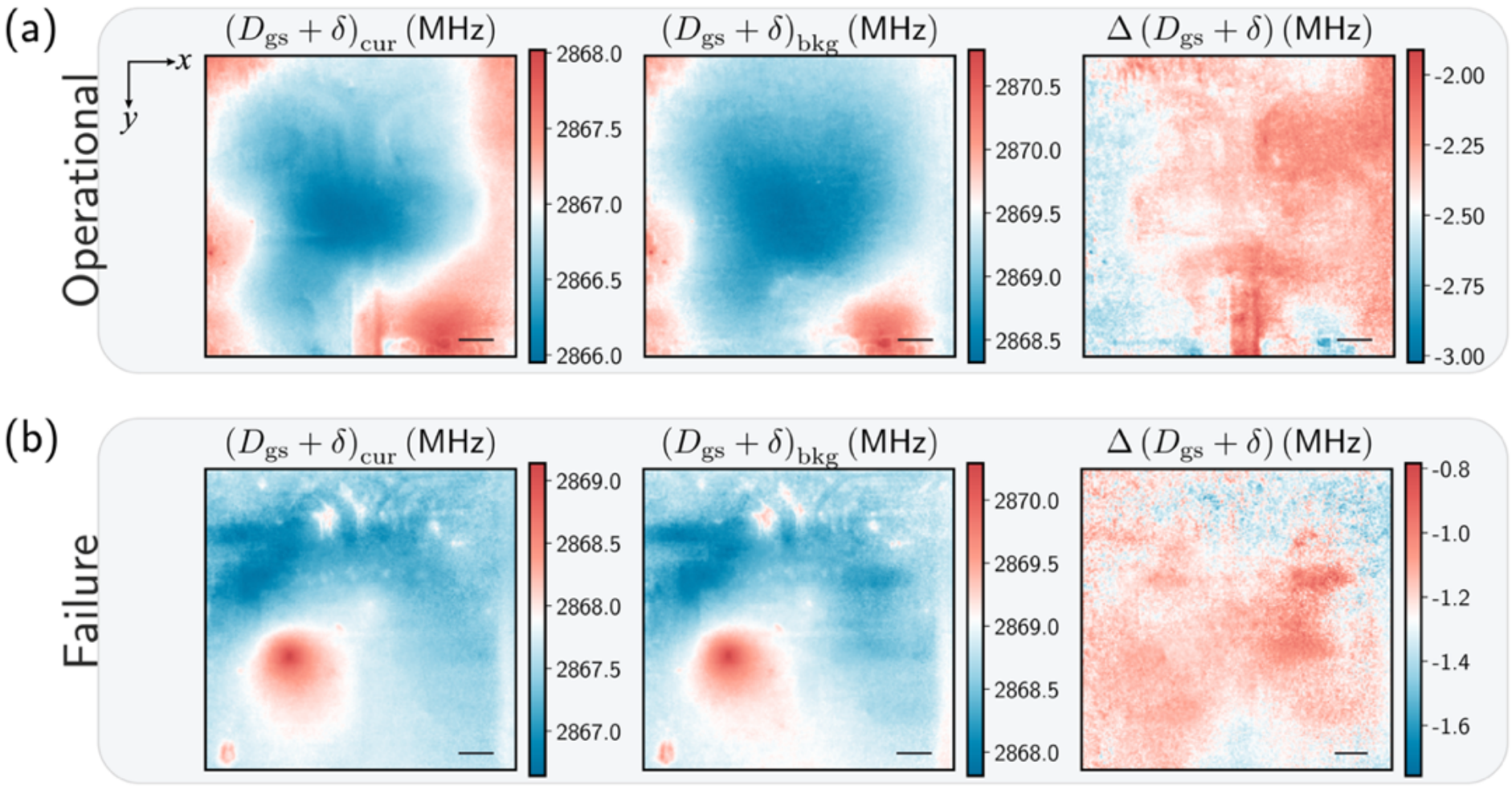

**Fig. S7. $D_{gs}+\delta$ parameters for the device 1 (operational) and the device 2 (failure). a)** $D_{gs}+\delta$ parameters for the operational case, with closed-circuit $(D_{gs}+\delta)_{cur}$, open-circuit $(D_{gs}+\delta)_{bkg}$ and the difference $\Delta(D_{gs}+\delta) = (D_{gs}+\delta)_{cur} - (D_{gs}+\delta)_{bkg}$. **b)** Same than in a) for the failure case.

In Fig. S7a (device 1, operational), the main effective parameter is the temperature. The temperature is higher in the centre of the FOV and colder at the edges. The temperature rise can be explained by the presence of transistors at the centre of the map (underneath the surface). The transistors produce heat, and contact with the diamond surface expels the heat. The difference between both maps shows an averaged rise of $\approx$ 30 K with closed circuits.

In Fig. S7b (device 2, failure), we can observe a singularity present in open and closed-circuit which could be explained by strain. Strain reduces the band-gap in semiconductors and then enhances leakage which can explain the loss of current amplitude observed in the main text (Fig. 2).

NV centres for which $\vec{B}_0$ is minimal along their quantisation axis (here NV centres along with the $\hat{u}_{NVC}$ and $\hat{u}_{NVD}$ axes) show a better sensitivity to the Stark effect [68,69]. This can enable us to seek electric field and strain effects simultaneously with the magnetic field and the ZFS, through the terms $\alpha, \beta$ as shown in Fig. S8 and Fig. S9.

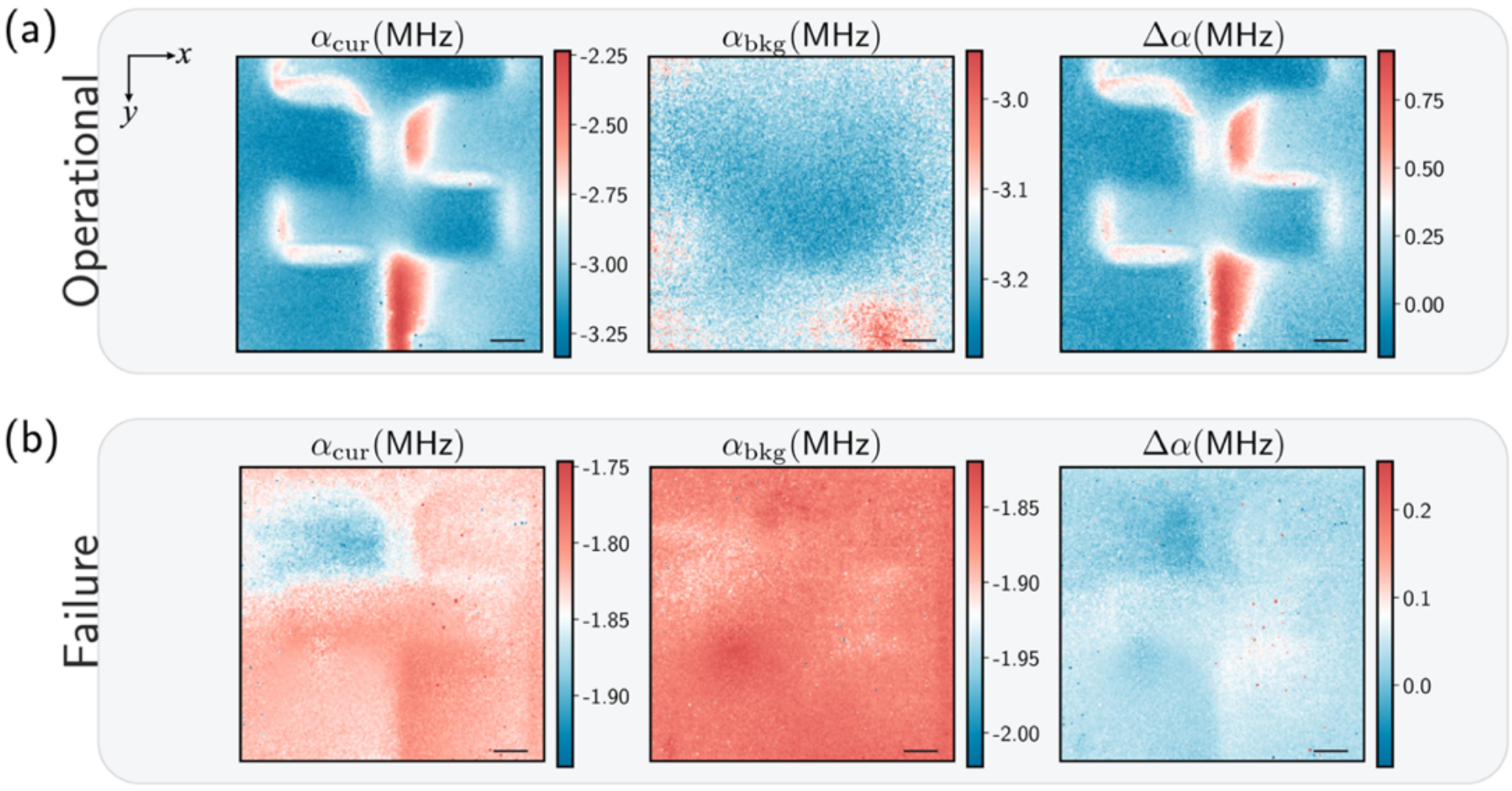

**Fig. S8. $\alpha$ parameter for the device 1 (operational) and the device 2 (failure). a)** $\alpha$ parameters for the operational case, with closed-circuit $\alpha_{\mathrm{cur}}$, open-circuit $\alpha_{\mathrm{bkg}}$ and the difference $\Delta\alpha = \alpha_{\mathrm{cur}} - \alpha_{\mathrm{bkg}}$. **b)** Same than in a) for the failure case.

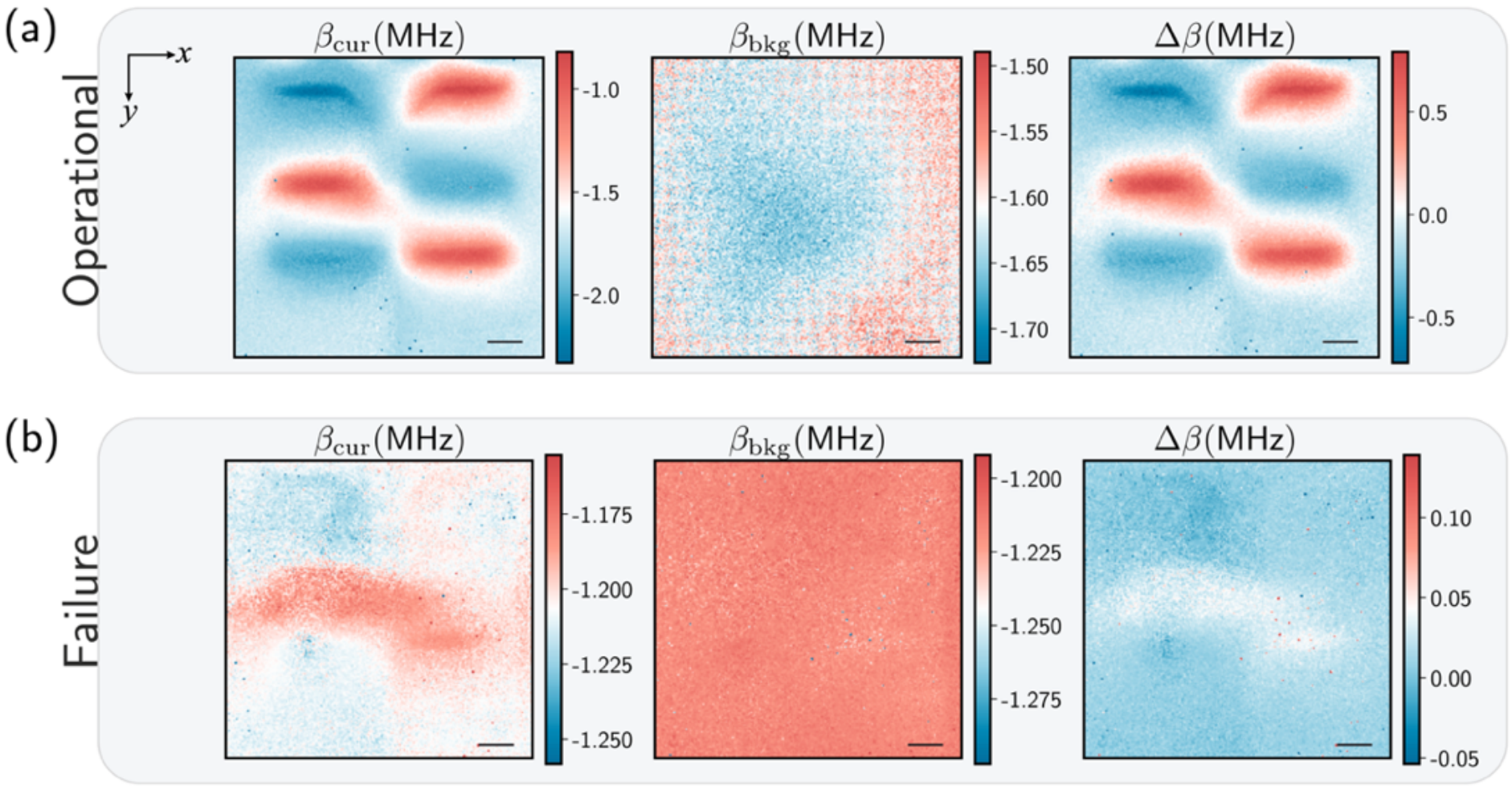

**Fig. S9. β parameter for the device 1 (operational) and the device 2 (failure). a)** $\beta$ parameters for the operational case, with closed-circuit $\beta_{\mathrm{cur}}$, open-circuit $\beta_{\mathrm{bkg}}$ and the difference $\Delta\beta = \beta_{\mathrm{cur}} - \beta_{\mathrm{bkg}}$. **b)** Same than in a) for the failure case.

In Fig. S8 and Fig. S9, when the current is off, the shifts can be unambiguously associated with internal strain inside the diamond. The values for strain yield similar ones to those previously reported for diamond plates [70]. Completing our experiment with spectroscopy measurements to confirm the internal strain of the diamond crystal would provide robust solutions to further conclude on electric effects. Finally, it is worth noting that considering $\alpha$ and $\beta$ as part of the free parameters in the optimisation process enhances the magnetic field maps' signal to noise ratio by approximately 1%. This value is relevant in Section S2, where the robustness of the current reconstruction relies on low noise magnetic field images.

Finally, the results suggest we can simultaneously monitor temperature rises, strain, and currents. This is promising for IC analysis, where peak temperatures reach about 400 K. In this regard, further evaluation and measurements to decorrelate all the effects will be further investigated in the future.

## S2. Current density reconstruction

To reconstruct the lateral current density $J_{xy}$, we need to solve the integral equation of the Biot-Savart law (Eq. (1) in the main text). We use the method based on deconvolution, using the Fourier Transform, also known as the standard magnetic inverse solution [39,40]. This method allows imaging of the current density distribution $\boldsymbol{J}(\mathbf{r}')$ at the source plane by considering the Biot-Savart law as a convolution, as follows:

$$\mathbf{B}(\mathbf{r}) = \frac{\mu_0}{4\pi} \boldsymbol{J}(\mathbf{r}') * \boldsymbol{G}(\mathbf{r} - \mathbf{r}'), \tag{S11}$$

where $\mathbf{B}(\mathbf{r})$ is the magnetic field detected at the observation position and $\boldsymbol{G}(\mathbf{r} - \mathbf{r}')$ is a green function describing the spreading of the magnetic field over the distance from the current source.

Using Fourier transform as described later in this section allow us to deconvolute Eq. (S11) and extract the $\boldsymbol{J}(\mathbf{r}')$ term. This method can partially compensate for the loss of spatial resolution due to field propagation. However, a major drawback is that the method amplifies the high-frequency components of the Fourier domain and, hence, noise. In order to compensate for the noise amplification effect, the procedure requires spatial filtering to dampen the high-frequency components. However, filtering tends, in turn, to smooth out the fine features of the image. Consequently, the magnetic field's signal-to-noise ratio must be optimised to obtain the most accurate current density mapping to limit the use of filters. In our case, this is achieved by considering a precise NV Hamiltonian when determining the magnetic field contribution as described in Section S1.

### S2.1. Two-dimensional current reconstruction

When considering a current confined in a 2D structure, laying in the *xy*-plane (i.e. $z = 0$), $\boldsymbol{J}$ can be redefined as a lineal current density expressed in units of A/m, which depends on the planar coordinates (*x,y*) only. The integral Eq.(1) from the main text becomes a 2D integral over (*x',y'*). With $\boldsymbol{J}^{\intercal} = \left(J_x \, J_y \; 0\right)$, we can therefore express the magnetic field components as:

$$B_x(x, y, z) = \frac{\mu_0 z}{4\pi} \int_{-\infty}^{+\infty} \int_{-\infty}^{+\infty} \frac{J_y(x', y')}{[(x - x')^2 + (y - y')^2 + z^2]^{3/2}} dx' dy' \tag{S12}$$

$$B_y(x, y, z) = \frac{\mu_0 z}{4\pi} \int_{-\infty}^{+\infty} \int_{-\infty}^{+\infty} \frac{-J_x(x', y')}{[(x - x')^2 + (y - y')^2 + z^2]^{3/2}} dx' dy \tag{S13}$$

$$B_z(x, y, z) = \frac{\mu_0}{4\pi} \int_{-\infty}^{+\infty} \int_{-\infty}^{+\infty} \frac{J_x(x', y')(y - y') - J_y(x', y')(x - x')}{[(x - x')^2 + (y - y')^2 + z^2]^{3/2}} dx' dy' \tag{S14}$$

If we now consider a 2D current in a slab delimited by the planes $\mathrm{z} = \mathrm{z}_1$ and $z = z_2$ (which can be determined by microscopy), the current density is expressed in units of A/m$^2$ and the equations can be rewritten as follows:

$$B_x(x,y,z) = \frac{\mu_0 z}{4\pi}\int_{-\infty}^{+\infty}\int_{-\infty}^{+\infty}\int_{z_1}^{z_2} \frac{J_y(x',y')}{[(x-x')^2+(y-y')^2+(z-z')^2]^{3/2}} dx'dy'dz' \tag{S15}$$

$$B_y(x,y,z) = \frac{\mu_0 z}{4\pi}\int_{-\infty}^{+\infty}\int_{-\infty}^{+\infty}\int_{z_1}^{z_2} \frac{-J_x(x',y')}{[(x-x')^2+(y-y')^2+(z-z')^2]^{3/2}} dx'dy'dz' \tag{S16}$$

$$B_z(x,y,z) = \frac{\mu_0}{4\pi}\int_{-\infty}^{+\infty}\int_{-\infty}^{+\infty}\int_{z_1}^{z_2} \frac{J_x(x',y')(y-y') - J_y(x',y')(x-x')}{[(x-x')^2+(y-y')^2+(z-z')^2]^{3/2}} dx'dy'dz' \tag{S17}$$

Additionally, we can add the continuity condition for the current, such as:

$$\nabla \cdot \boldsymbol{J} = 0 \tag{S18}$$

Using the Fourier transform defined as:

$$f(k_x,k_y,z) = \int_{-\infty}^{+\infty}\int_{-\infty}^{+\infty} F(x,y,z)e^{k_x x + k_y y}\, dxdy \tag{S19}$$

we can rewrite the set of equations in the Fourier space $\{\mathrm{k_x}, \mathrm{k_y}\}$, such as:

$$\begin{pmatrix} b_x \\ b_y \\ b_z \\ 0 \end{pmatrix} = \frac{\mu_0}{4\pi}\begin{pmatrix} 0 & g_z \\ -g_z & 0 \\ g_y & -g_x \\ ik_x & ik_y \end{pmatrix} \cdot \begin{pmatrix} j_x \\ j_y \end{pmatrix} \tag{S20}$$

Where $\{b_x, b_y, b_z\}$, $\{j_x, j_y, 0\}$, $\{g_x, g_y, g_z\}$ are respectively the Fourier components of the magnetic field $\boldsymbol{B}$, the current density $\boldsymbol{J}$, and the green function $G$.

With two unknowns and four equations, we have an overdetermined system. Eq. (S20) can be solved as it is but it is computationally more expensive to solve an overdetermined system compared to a determined system. Additionnally, the use of $b_z$ introduces a singularity and tends to produce artefacts (see discussion later in this section). Thus, we reduce the system to a determined system, such as:

$$\begin{pmatrix} b_x \\ b_y \end{pmatrix} = \frac{\mu_0}{4\pi}\begin{pmatrix} 0 & g_z \\ -g_z & 0 \end{pmatrix} \cdot \begin{pmatrix} j_x \\ j_y \end{pmatrix} \tag{S21}$$

To solve this system where the Fourier components $\{j_x, j_y, 0\}$ are the unknown parameters, we use a QR decomposition method [71]. In addition, we use a Hanning window on $j_x$ and $j_y$ to filter high-frequency components smoothly. Finally, we apply an inverse Fourier transform to the returned parameters $j_x$ and $j_y$ and apply total variation denoising using the Chambolle algorithm [72] on the data.

As discussed in [73], the use of the lateral components $B_x$ and $B_y$ shows better performances for current reconstruction than using the vertical component $B_z$. One reason is that $B_z$ introduce a singularity in the deconvolution process since the components $k_x = k_y = 0$ are undistiguishable. Besides, $B_z$ has an intrinsic longer-range behaviour (i.e. larger lateral spreading) compared to the lateral components $B_x$ and $B_y$, inducing artefacts in the current reconstruction procedure. Additionally, the distinct separation between $B_x$ and $B_y$ offers better signal readability than seeking $B_z$ only, as with SQUIDs. This ability to distinguish $B_x$ from $B_y$ is crucial in multi-layered chips where orthogonal signals can overlap at the observation position.

### S2.2. Three-dimensional current reconstruction

Finally, knowing $B_x, B_y$ grants us access to a determined system when the three-dimensional components of the current density are unknown parameters. The equations rewrite as follows:

$$B_x(x,y,z) = \frac{\mu_0}{4\pi}\int_{-\infty}^{+\infty}\int_{-\infty}^{+\infty}\int_{z_1}^{z_2}\frac{J_y(x',y',z')\Delta z - J_z(x',y',z')\Delta y}{[\Delta x^2 + \Delta y^2 + \Delta z^2]^{3/2}}dx'dy'dz' \quad \text{(S22)}$$

$$B_y(x,y,z) = \frac{\mu_0}{4\pi}\int_{-\infty}^{+\infty}\int_{-\infty}^{+\infty}\int_{z_1}^{z_2}\frac{J_z(x',y',z')\Delta x - J_x(x',y',z')\Delta z}{[\Delta x^2 + \Delta y^2 + \Delta z^2]^{3/2}}dx'dy'dz' \quad \text{(S23)}$$

$$B_z(x,y,z) = \frac{\mu_0}{4\pi}\int_{-\infty}^{+\infty}\int_{-\infty}^{+\infty}\int_{z_1}^{z_2}\frac{J_x(x',y',z')\Delta y - J_y(x',y',z')\Delta x}{[\Delta x^2 + \Delta y^2 + \Delta z^2]^{3/2}}dx'dy'dz' \quad \text{(S24)}$$

where:

$$\Delta x = (x - x'),$$

$$\Delta y = (y - y'),$$

$$\Delta z = (z - z').$$

In the Fourier space, this translate to:

$$\begin{pmatrix} b_x \\ b_y \\ b_z \\ 0 \end{pmatrix} = \frac{\mu_0}{4\pi}\begin{pmatrix} 0 & g_z & -i\frac{g_z k_y}{k} \\ -g_z & 0 & i\frac{g_z k_x}{k} \\ i\frac{g_z k_y}{k} & -i\frac{g_z k_x}{k} & 0 \\ -i\frac{k_x}{k} & -i\frac{k_y}{k} & \frac{\partial j_z}{\partial z} \end{pmatrix} \cdot \begin{pmatrix} j_x \\ j_y \\ j_z \end{pmatrix} \quad \text{(S25)}$$

Again, we leave the contribution from $b_z$ out, reducing the system as follows:

$$\begin{pmatrix} b_x \\ b_y \\ 0 \end{pmatrix} = \frac{\mu_0}{4\pi} \begin{pmatrix} 0 & g_z & -i\frac{g_z k_y}{k} \\ -g_z & 0 & i\frac{g_z k_x}{k} \\ -i\frac{k_x}{k} & -i\frac{k_y}{k} & \frac{\partial j_z}{\partial z} \end{pmatrix} \cdot \begin{pmatrix} j_x \\ j_y \\ j_z \end{pmatrix} \tag{S26}$$

It is important to note that $B_x$ and $B_y$ must be first described in a right-handed coordinate system to solve Eq. (S26). The components $J_x$,$J_y$,$J_z$ are the current components at the source plane. In order to solve the current distribution over the entire device, each layer should be considered, and the current in the vias can be determined from $J_z$ in each layer and the continuity condition of the current. In the main text, we only focus on the outer layer of a thickness of about 3 µm since the lateral spreading in the deeper layer is too prominent to reconstruct the current precisely without adding knowledge about the layout. As for the 2D case, we use a QR decomposition method to solve the system and apply a Hanning window on $j_x$, $j_y$ and $j_z$ to filter high-frequency components. Finally, we apply an inverse Fourier transform to the returned parameters and apply the Chambolle denoising algorithm, resulting in the $J_x$,$J_y$,$J_z$ maps shown in Fig. 5, a-c, of the main text. The resulting map $|\mathrm{J_{tot}}| = \sqrt{\mathrm{J_x^2} + \mathrm{J_y^2} + \mathrm{J_z^2}}$ is shown in Fig. S10. Using the information of each vectorial component of the current density $J_x$,$J_y$,$J_z$, the flux can be determined as pictured in Fig. 5d of the main text. However, to resolve the complete path as shown in Fig. S10, further analysis is needed: firstly, we need to resolve the current's contribution in the vertical direction as discussed in Section S3, and secondly, we need to resolve counter-propagative fields as discussed in Section S4.

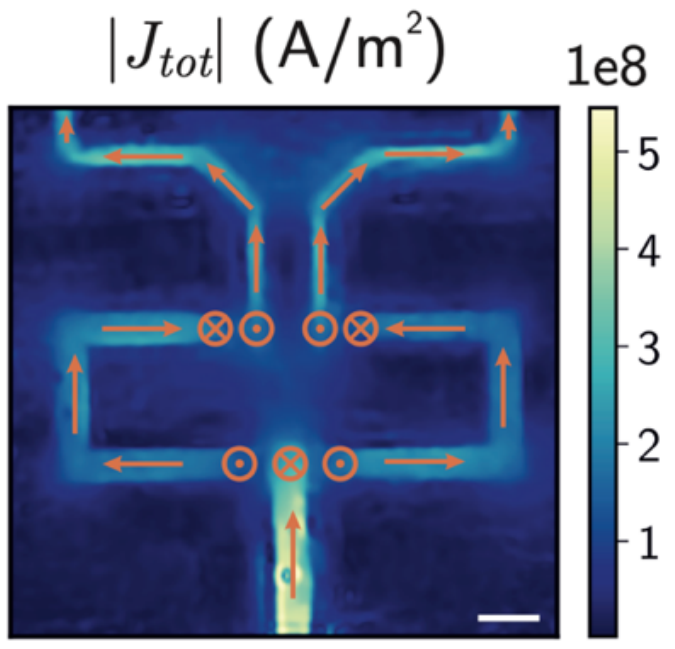

**Fig. S10. Total current density $|J_{tot}|$.** Orange arrows indicate the current's direction in the outer layer. The scale bar is 10 µm wide.

## S3. Current source localisation in a multi-layered device

The magnetic field sensed by a layer of NV centres shown in Fig. 2, c and d (main text), combines the contribution of each active layer. Since the measurements are perfomed outside the device to keep it undamaged, all the contributions over $z$ are merged and this section aims at retriving the source's localisation.

The Biot-Savart law (Eq. (1) in the main text) describes the evolution of the magnetic field produced by a current source over the distance of observation. Therefore, we can localise currents in space by exploring the source-sensor distance dependence of the law. Linecuts in $B_x$ at positions $y = y_A, y = y_B$ for the device 1 are shown in Fig. 3a and linecuts at positions $y = y_{A'}, y = y_{B'}$ for the device 2 are shown in Fig. 3b. Finally, linecuts in $B_y$ at positions $x = x_c$ for the device 1 and at $x = x_{c'}$ for the device 2 are shown in Fig. S11.

As mentioned in the main text, we fit the curves using the Biot-Savart, using the infinite wire approximation (Eq. (2) in the main text). Mainly, the fit informs us about the total current amplitude in the leads and the localisation of the current sources with respect to the observation position in cartesian coordinates $\{x, y, z\}$. At first, we consider parameters $I_{xy}, r_{\text{wire}}, \Delta z$, and $o$ as free parameters and the fitting suggest two layers depths: the first one is at 4.5(5) µm from the surface of the diamond and the second at 8.5(8) µm. We proceed to the fitting a second time, allowing the two above-mentioned vertical positions only, resulting in the fitting in Fig. 3, a and b (main text). We proceed in the same way for the fitting in Fig. S11.

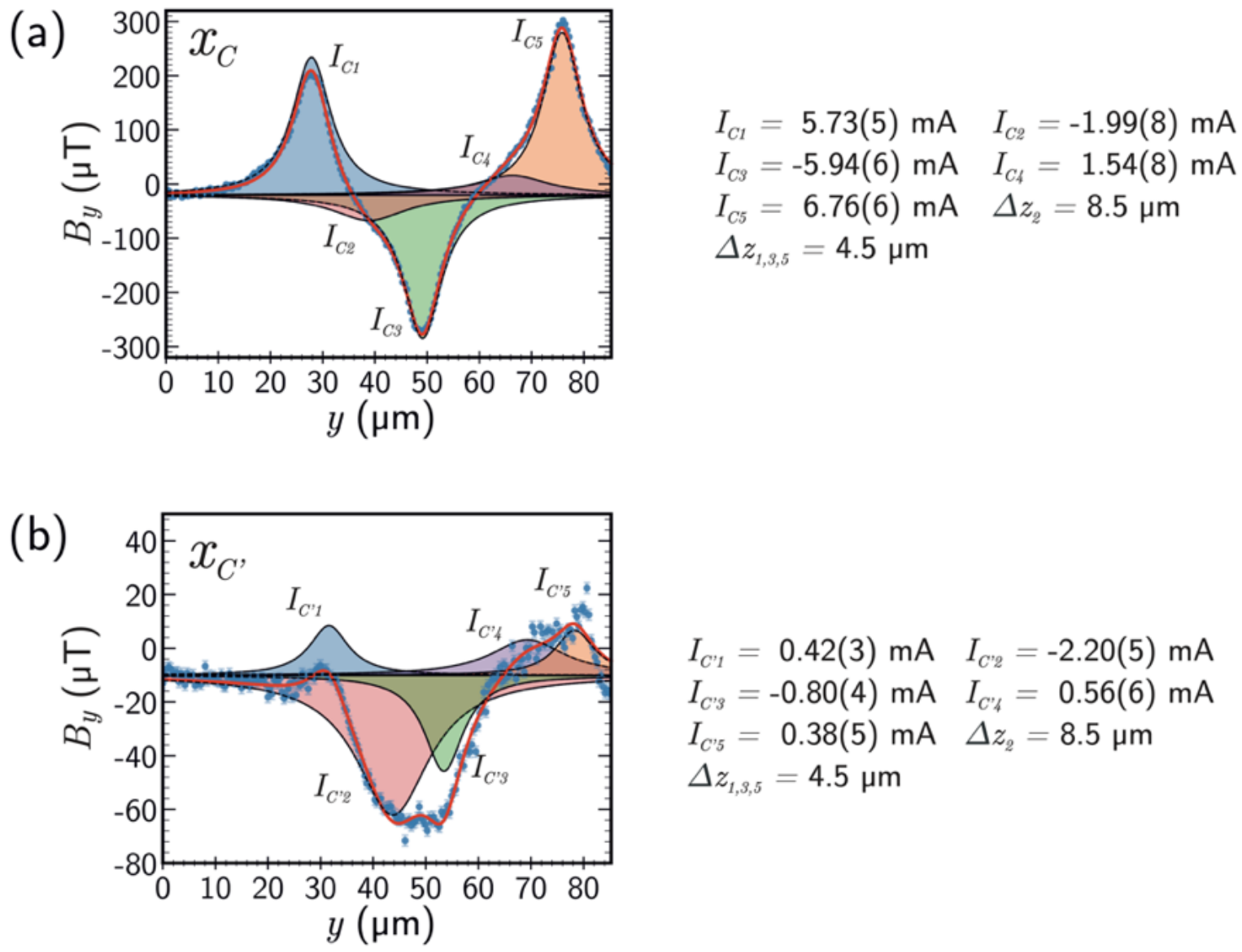

**Fig. S11. Linecuts of $B_y$. a)** Linecut along $y$ at the position $x = x_C$ for the device 1 (operational). The curve was fitted with a 5 wires model (red solid line) and the returned parameters are shown on the right-hand side of the curve. **b)** Same as a) for the device 2 (defective).

Feeding the fitting algorithm with layout information will significantly improve the precision of the fit. For instance, the precision of the vertical source-sensor distance would be limited by the NV centres depth distribution, spanning about 20 nm in our case.

## S4. Simulation of Oersted fields generated in a multi-layered device

In Fig. 4 of the main text, we have shown a simulation of Oersted fields generated in a multi-layered device comprising through-silicon vias (TSVs). The complexity in non-destructive imaging of such devices' activity is that the signal blurs with distance, and signals arising from distinct parts can merge and create complex patterns. The ability to recognise patterns and differentiate them is a key feature that needs further development in current reconstruction methods. Here we show how the contributions from different parameters differ and hence can be used with pattern recognition algorithms.

Fig. S12 shows the simulated Oersted fields created by each AL and the TSVs at the sensor position.

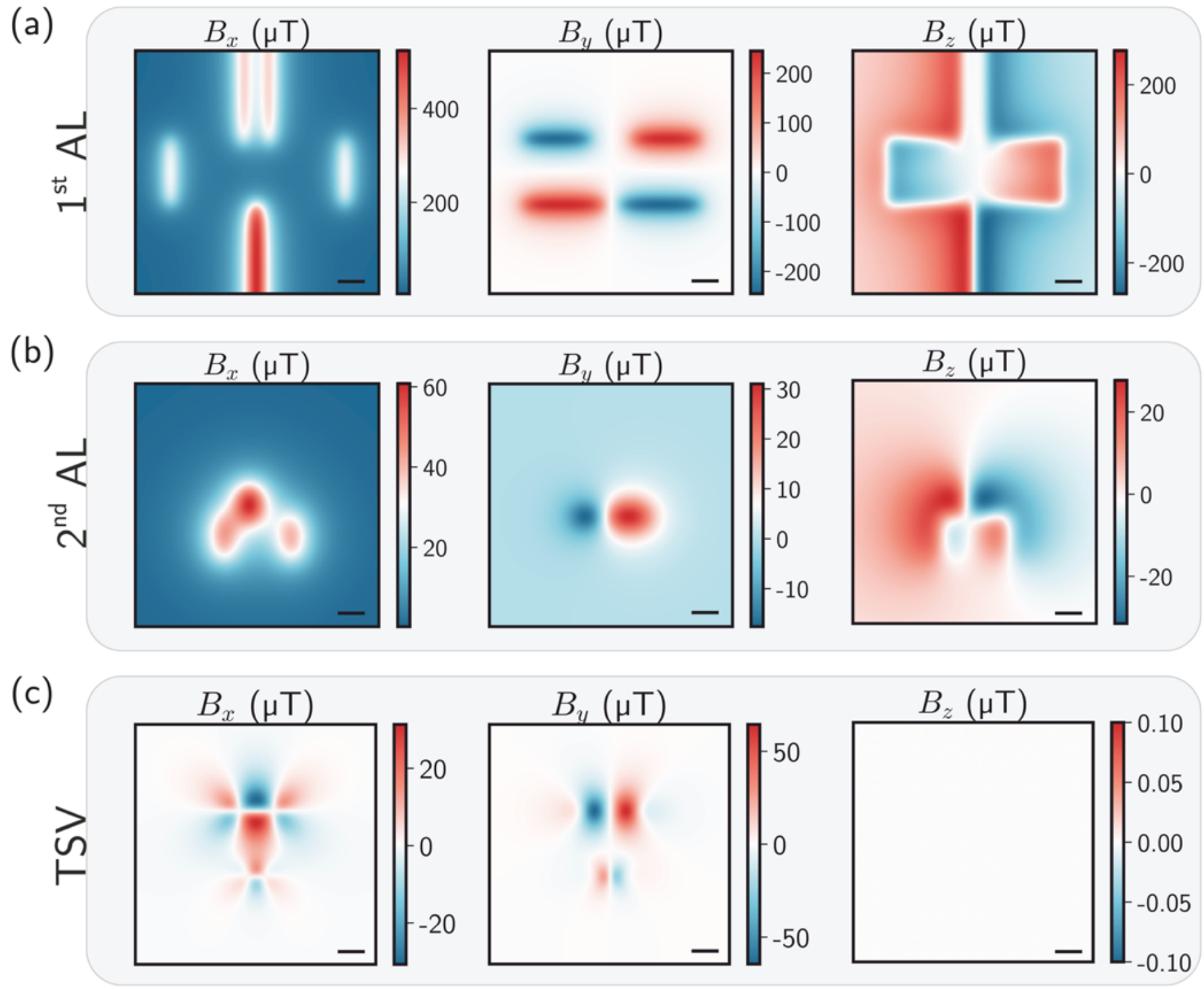

**Fig. S12. Top view of Oersted fields generated by the active element of the simulated structure. a)** $B_x, B_y, B_z$ patterns generated by the 1st AL at the sensors position. **b)** $B_x, B_y, B_z$ patterns generated by the 2nd AL at the sensors position. **c)** $B_x, B_y, B_z$ patterns generated by the TSVs at the sensors position. Scale bars are 10 μm wide.

### S4.1. TSVs

An essential aspect of this simulation is the Oersted fields generated by the TSVs since they produce no signal in $B_z$, as the current flows only along the vertical axis $z$. Accordingly, methods based on measuring the out-of-plane component of the magnetic field $B_z$ alone (e.g. standard SQUID-based microscopes) are not sufficient to image activity produced by vertical components. Hence, it is worth noting that NV-based imaging offers, as of today, the best alternative to imaging TSVs' activity with high spatial resolution.

Several factors can affect the magnetic pattern produced by the TSVs: the number of wires contributing to the pattern, the length $l_z$ of each wire, the current amplitude $I$ in a wire and its direction, and the distance of observation. In this section, we investigate the effect of each of these parameters on the simulated structure.

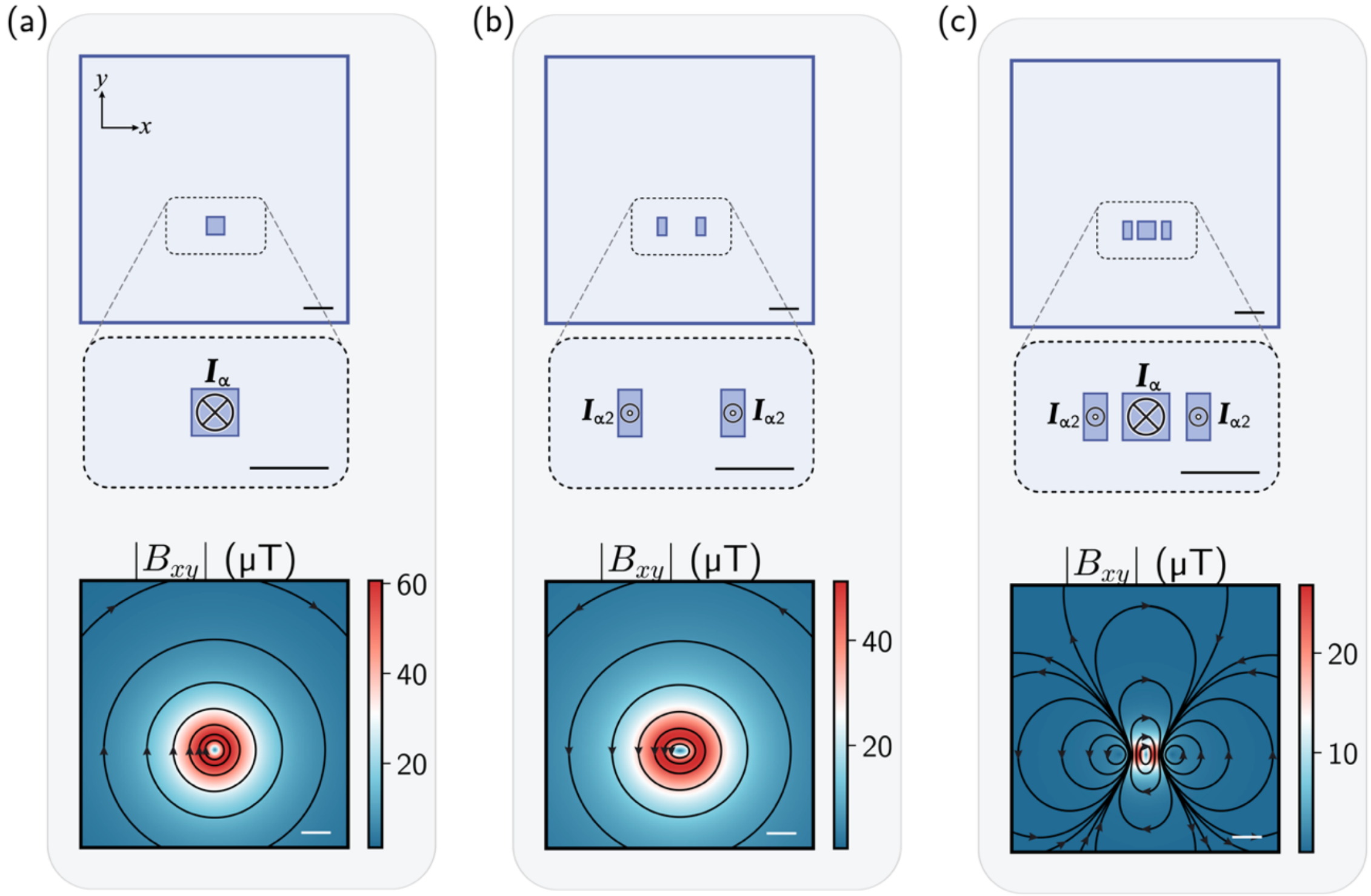

**Fig. S13. Top view of the Oersted fields generated by the TSVs. a)** *Top*: Scheme of the simulated structure with a zoom-in around the active lead where a current of amplitude $I_\alpha$ flows from the 1st AL to the bottom layer of the chip. *Bottom*: Corresponding $|B_{xy}|$ pattern generated by the simulated TSV at the position of the sensors. **b)** *Top*: Scheme of the simulated structure with a zoom-in around the active leads where a current of amplitude $I_{\alpha2} = I_\alpha/2$ flows from the bottom layer of the chip to the 1st AL. *Bottom*: Corresponding $|B_{xy}|$ pattern generated by the simulated TSVs at the position of the sensors. **c)** *Top*: Scheme of the simulated structure with a zoom-in around the active leads combining current distributions from (a) and (b). *Bottom*: Corresponding $|B_{xy}|$ pattern generated by the simulated TSVs at the position of the sensors. Solid black lines with arrows represent the magnetic field flux. Scale bars are 10 µm wide.

In Fig. S13, we show the magnetic field's amplitude patterns $|B_{xy}| = \sqrt{B_x^2 + B_y^2}$ and flux lines produced by each TSV in the simulated circuit.

First, we separate the contribution from counter-propagating elements, resulting in the Fig. S13, a and b. The pattern produced by a single wire is simple: the magnetic field is null at the centre of the wire and has a radial symmetry around the axis of the wire. Downstream currents produce clockwise flux lines, while upstream currents produce counter-clockwise ones. Compared to the single wire's pattern, the pattern created by two proximate co-propagative wires is slightly different. The magnetic field is null where the flux lines from each wire cancel out, and the magnetic amplitude pattern depends on the distance between the two wires. Finally, when we gather the contribution from all the wires, which includes counter-propagative currents, the pattern becomes more complex, as shown in Fig. S13c. The overall magnetic field is reduced, and it becomes challenging to determine the number of wires from the magnetic amplitude pattern alone. However, observing the magnetic flux lines gives additional information: we can observe three significant loops corresponding to the number of wires. Therefore, observing magnetic flux lines added to the amplitude pattern is a powerful tool for determining the number of contributive wires in magnetic field analysis.

In Fig. S14, we now consider the impact induced by the total amplitude $I$ in a single TSV on the resulting magnetic field pattern. We vary the parameter $I$ while keeping all the other parameters constant and observe a linear relationship between the total current $I$ in a wire and the Oersted field amplitude. The higher the total current $I$ in a wire, the higher is the Oersted field amplitude and the wider is the resulting magnetic field pattern.

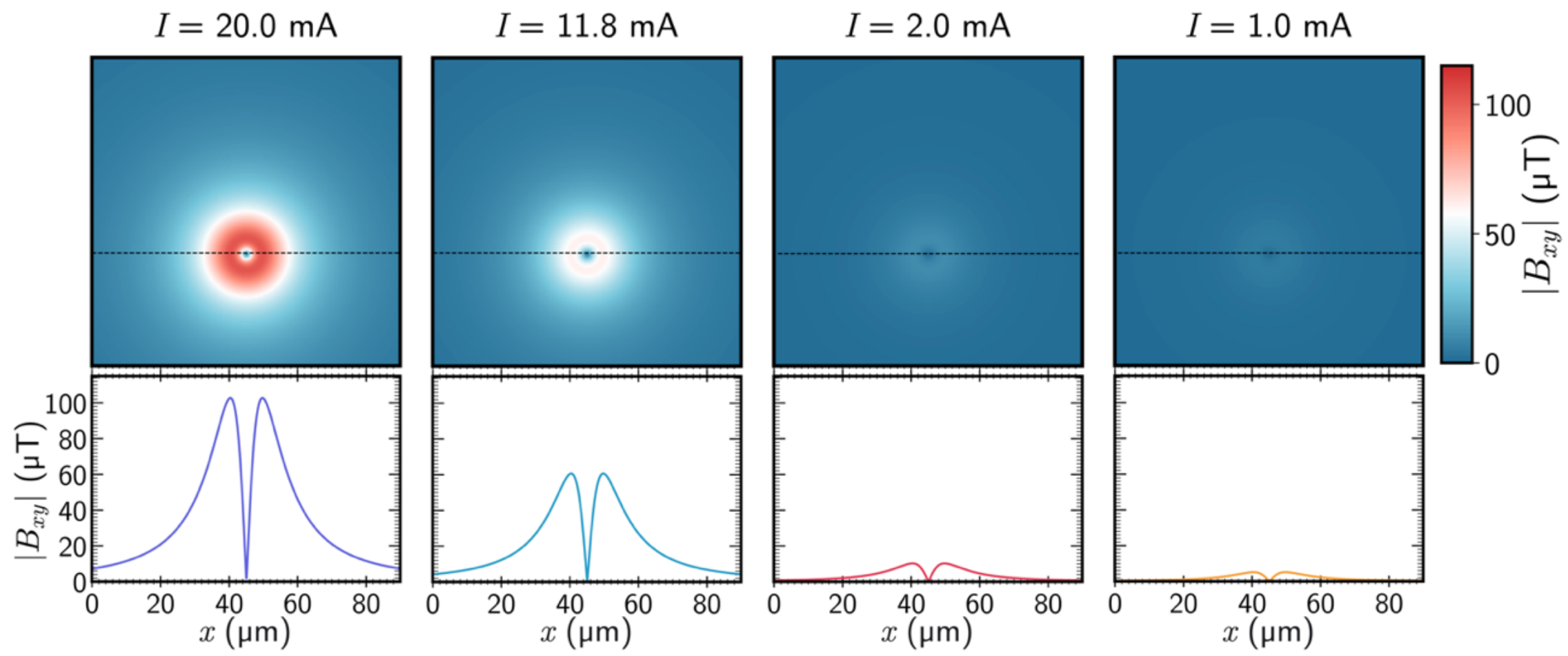

**Fig. S14. B-field pattern for a vertical wire with different current amplitude $I$.** We observe the field at a distance $\Delta z = 4.5$ µm for a single wire of length $l_z = 7.7$ µm. *Top*: Magnetic field patterns. *Bottom*: Magnetic field profile taken along the dashed black line on each magnetic field map. The second panel from the left shows the parameters used for the central TSV in the simulated structure.

Yet, another parameter can increase the total amplitude $|B_{xy}|$: the length $l_z$ of a single TSV. Thus, we consider in Fig. S15, the impact the length $l_z$ of a single TSV has on the magnetic field pattern. The Oersted field produced by a wire of length $l_z$ is the result of infinitesimals' contributions integrated over $l_z$. As a result, the longer the length $l_z$ of a wire, the higher is the Oersted field amplitude. Again, the resulting magnetic field pattern widens with $l_z$.

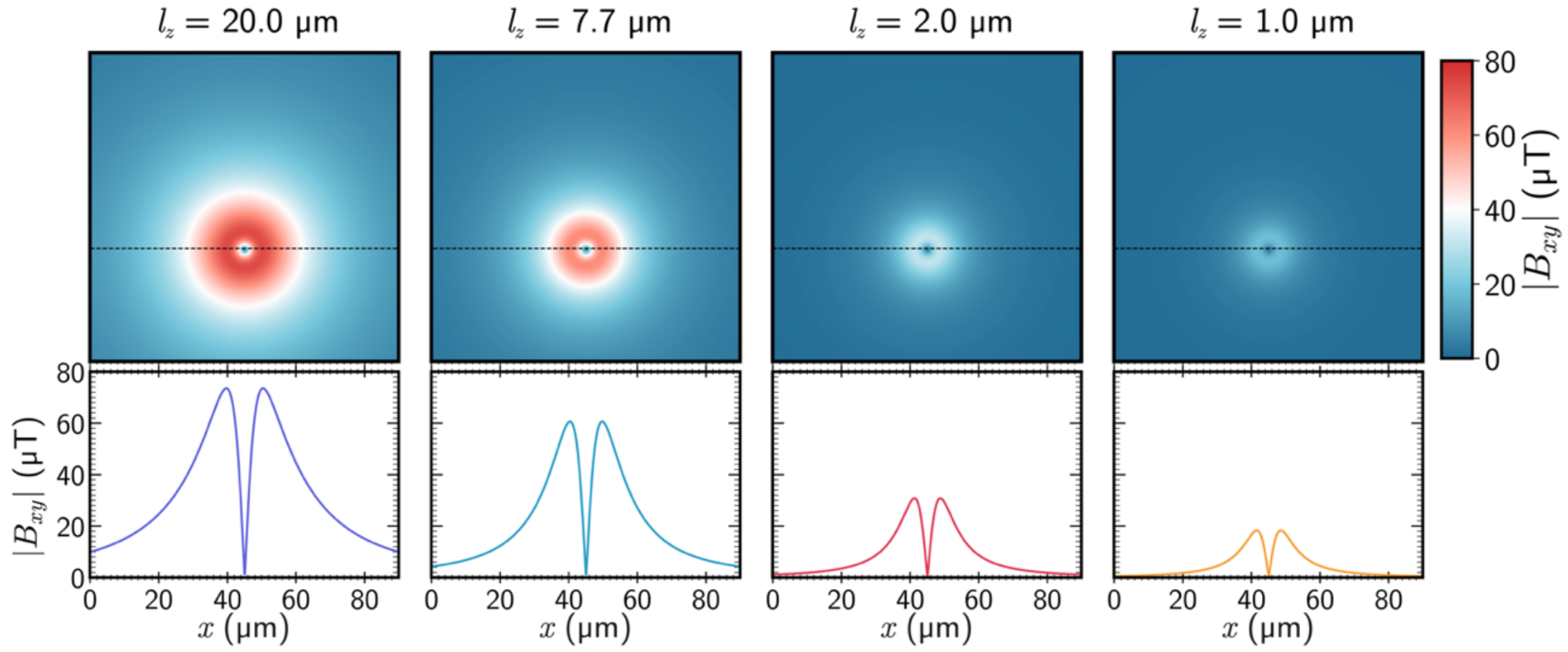

**Fig. S15. B-field pattern for a vertical wire with different length $l_z$.** We observe the field at a distance $\Delta z = 4.5$ μm for a single wire of length $l_z$. The total current amplitude is $I_\alpha = 11.8$ mA. *Top*: Magnetic field patterns. *Bottom*: Magnetic field profile taken along the dashed black line on each magnetic field map. The second panel from the left shows the parameters used for the TSV connected to the main lead in the simulated structure.

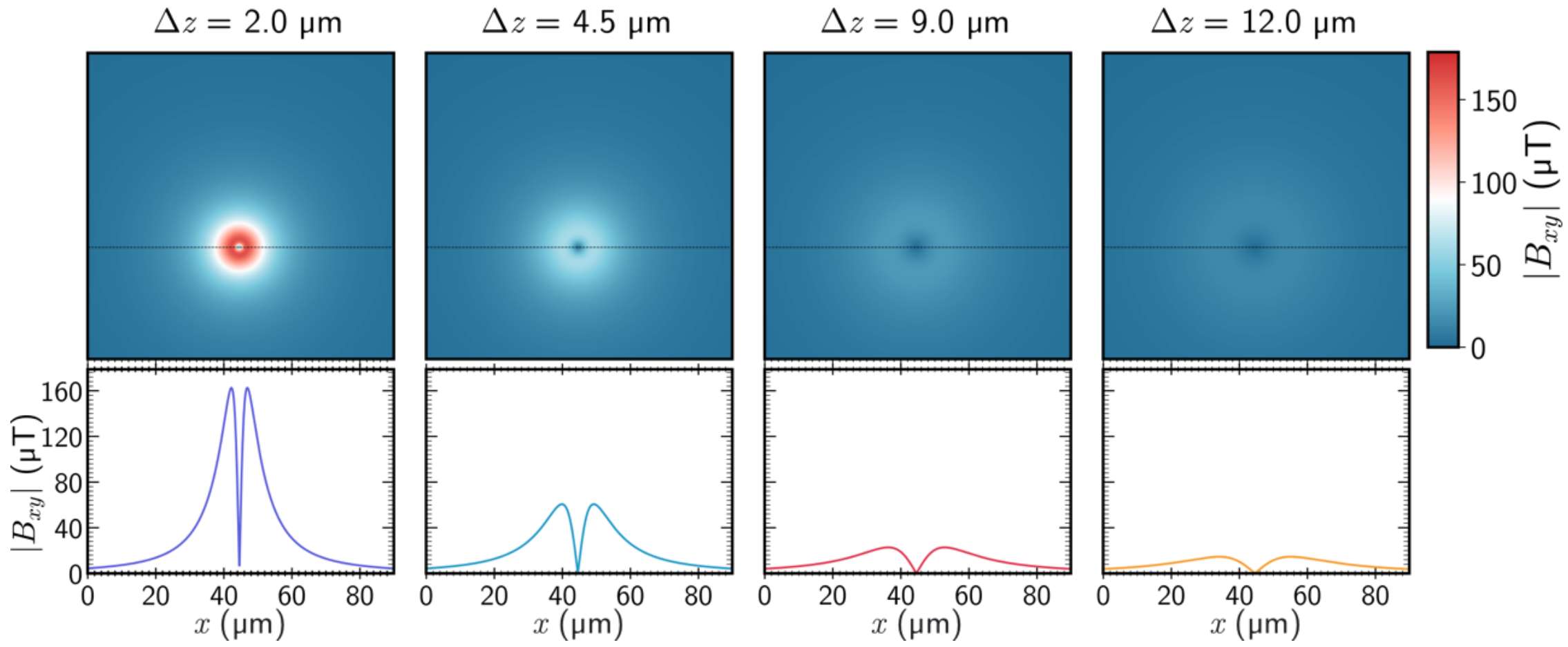

**Fig. S16. B-field pattern for a vertical wire observed at different distances $\Delta z$.** We observe the field for a single wire of length $l_z = 7.7$ μm and with a total current amplitude $I_\alpha = 11.8$ mA. *Top*: Magnetic field patterns. *Bottom*: Magnetic field profile taken along the dashed black line on each magnetic field map. The second panel from the left shows the parameters used for the TSV connected to the main lead in the simulated structure.

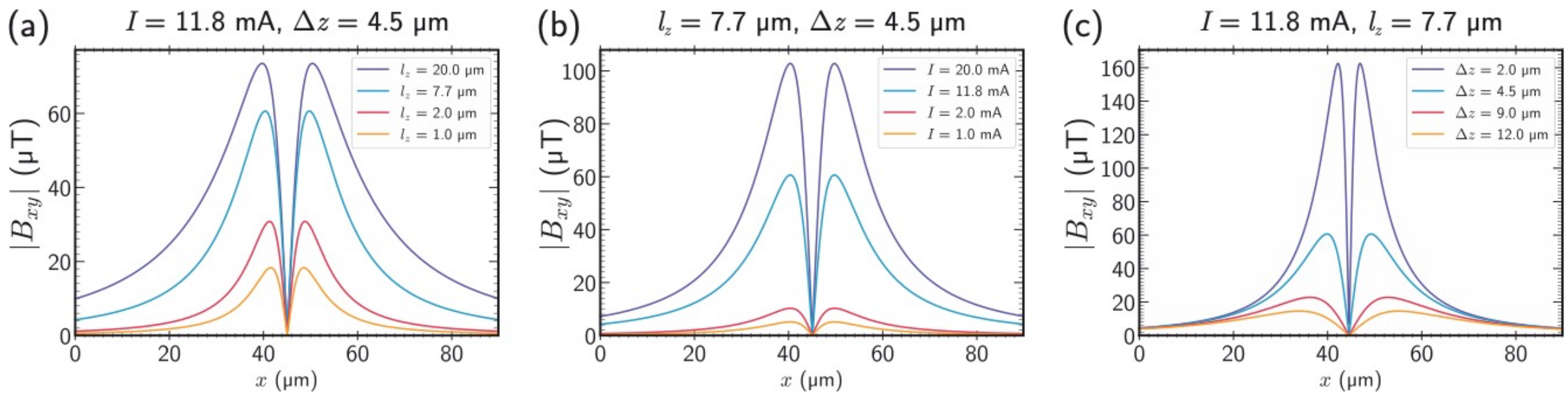

**Fig. S17. B-field profiles for a vertical wire with different parameters. a)** We observe the field at a distance $z = 4.5$ µm for a single wire of length $l_z$. The total current amplitude is $I_\alpha = 11.8$ mA. **b)** We observe the field at a distance $\Delta z = 4.5$ µm for a single wire of length $l_z = 7.7$ µm.

Finally, depending on the distance of observation (i.e. the depth of the vias), the magnetic field pattern can also widen and drastically flatten out over a few micrometres only. The effect of distance is simulated and shown in Fig. S16.

Overall, we can see in Fig. S17 the impact of the parameters *I*, $l_z$ and $\Delta z$ scales differently. Hence the dependence of $|B_{xy}|$ with those parameters can be differentiated. However, in complex structures where many active layers overlap, it can be difficult to separate the signals.
To this end, we can use the advange of knowing the three components of the magnetic field and use the information given by $(B_{xy} - B_z)$ to isolate the contributions from TSVs only. Exploring these features for developing pattern recognition algorithms using machine learning is extremely promising to analyse complex chip designs.

### S4.2. Overlapping wires

Another aspect to consider in multi-layered devices is the contribution of overlapping wires. In this section, we discuss several case scenarios.

Firstly, we consider two overlapping wires with different dimensions, as shown in Fig. S18a. A current $I_\alpha$ runs along the *y*-axis in a wire in the 1$^{st}$ AL and in a second wire in the 2$^{nd}$ AL, generating the Oersted fields shown in Fig. S18b. Since the magnetic pattern spreads over the distance *z*, there is a clear pattern's distortion with a local rise in intensity due to the wire in the 2$^{nd}$ AL. A complementary way to assess the presence of the underneath wire is to look at the profile of $B_x$ along the current's direction (i.e. the *y*-axis) as shown in Fig. S18c. The Oersted field generated by the wire in the 2$^{nd}$ AL adds a bell-shaped pattern to the otherwise almost flat $B_x$ profile. Hence, local rises in the Oersted field maps and profiles along the current's direction can help identify the wire's contributions from different layers. It is important to note that the inspection of $B_x$ along the *y*-axis is only possible because the NV centres resolve the three components $B_x, B_y, B_z$ separately, contrary to sensors acquiring $B_z$ only.

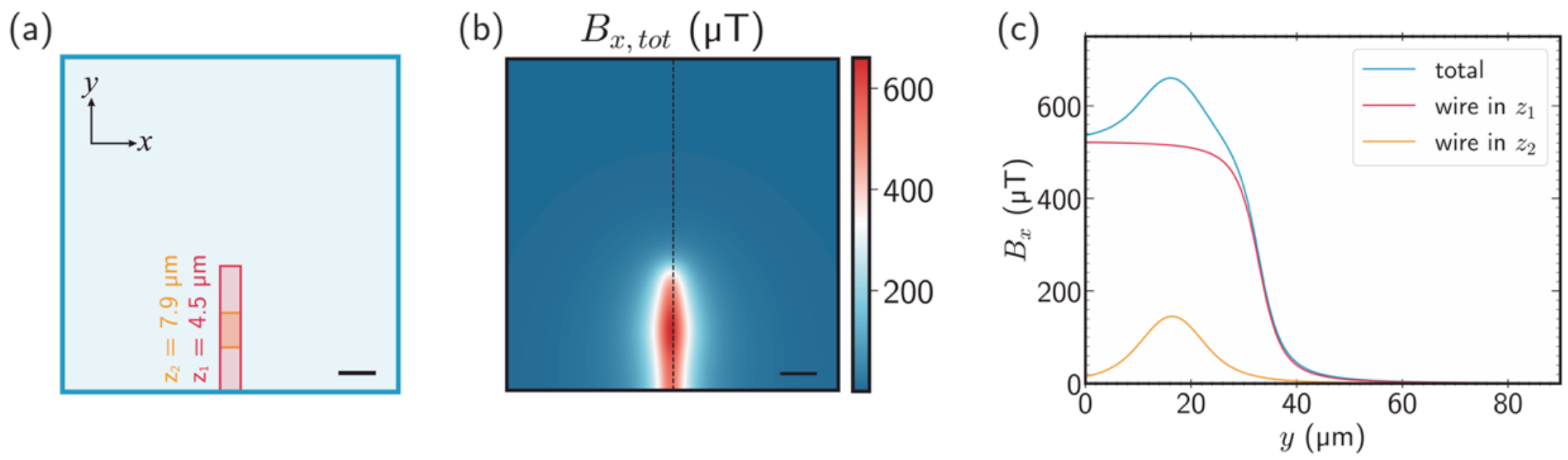

**Fig. S18. Top view of Oersted fields generated by overlapping wires of different dimensions. a)** Schematic top view of the overlapping wires. **b)** $B_{x,\mathrm{tot}}$ pattern generated by the combined contributions of a current $I_\alpha$ in the leads in the 1st AL and 2nd AL, at the sensors position. **c)** $B_x$ profile along the dashed line in (b). Blue line represents the combined contribution from both wires; red line shows the contribution from the wire in the 1st AL (i.e. $z = z_1$) and orange line shows the contribution from the wire in the 2nd AL ($z = z_2$). Scale bars in (a)-(b) are 10 µm wide.

Secondly, we consider overlapping wires of identical dimensions with two scenarios: wires with partial overlap as shown in Fig. S19a and wires perfectly overlapping as shown in Fig. S20a. Again, a current of amplitude $I_\alpha$ runs along with the *y*-axis in a wire in the 1st AL and in a second wire in the 2nd AL, generating the Oersted fields shown in Fig. S19, b-d, and Fig. S20, b-d. A linecut along the orthogonal axis of the wire (i.e., *x*-axis) is shown in Fig. S21. In the case of partially overlapping wires, the lateral shift between the wires results in an asymmetrical magnetic pattern. This asymmetry underlines the contribution of different wires and helps determine the wires' contributions. In the case of fully overlapping wires, finding evidence of multi-wire contributions is more difficult. The depths of the wires need to be known such that the wings of the magnetic-field profile (vanishing sides) can help determine the wires' contribution.

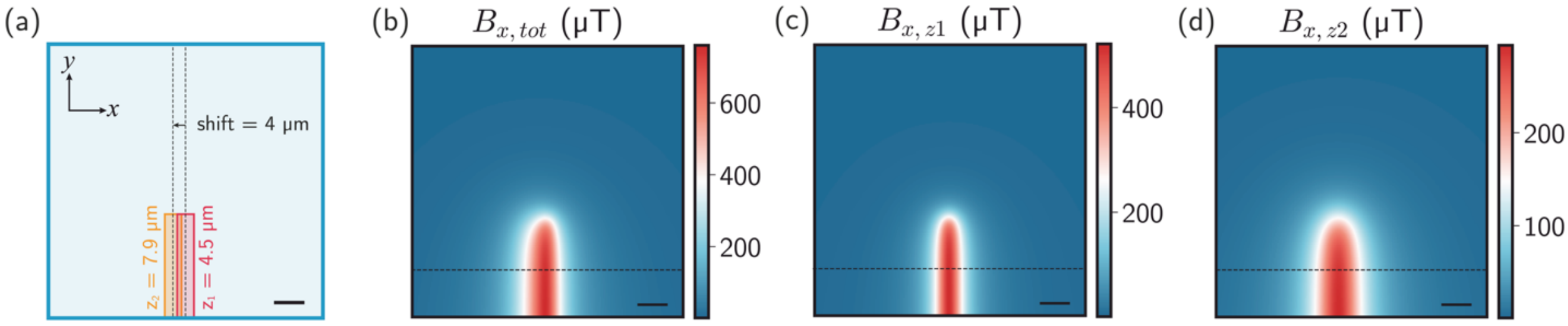

**Fig. S19. Top view of Oersted fields generated by partially overlapping wires. a)** Schematic view of the wires. **b)** $B_x$ pattern generated by the combined contributions of the main leads in the 1st and 2nd ALs, at the sensors position. **c)** $B_x$ pattern generated by the contribution of the main lead in the 1st AL, at the sensors position. **d)** $B_x$ pattern generated by the contribution of the main lead in the 2nd AL, at the sensors position. Scale bars are 10 µm wide.

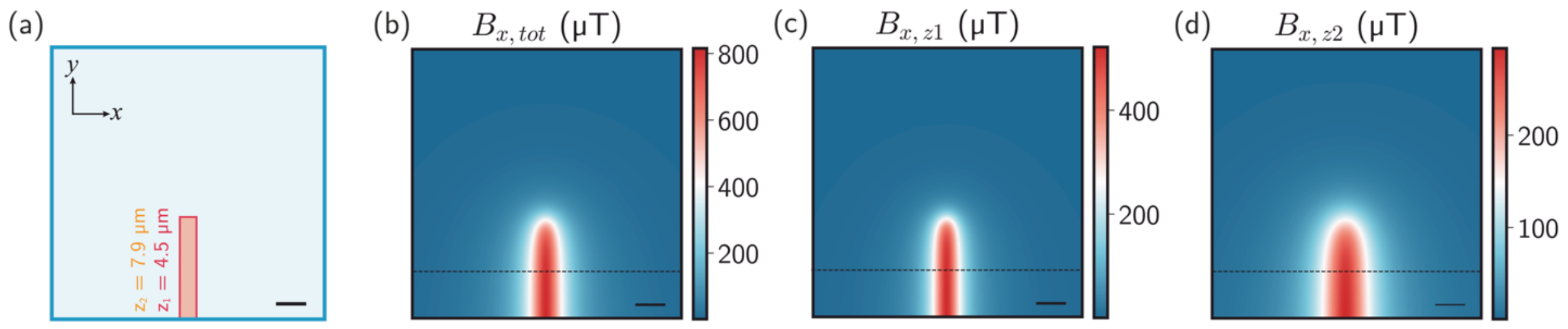

**Fig. S20. Top view of Oersted fields generated by overlapping wires. a)** Schematic view of the wires. **b)** $B_x$ pattern generated by the combined contributions of the main leads in the 1st and 2nd ALs, at the sensors position. **c)** $B_x$ pattern generated by the contribution of the main lead in the 1st AL, at the sensors position. **d)** $B_x$ pattern generated by the contribution of the main lead in the 2nd AL, at the sensors position. Scale bars are 10 µm wide.

In both cases, fitting a single profile line with no prior information is insufficient to determine the number of wires, depth, and current contribution. If the Oersted field is imaged through a larger area where the paths differentiate, then optimisation of the fitting procedure over the circuit can lead to robust results for devices with a simple design (i.e. single wires are distinguishable on some parts of the circuit). Otherwise, the device's geometry must be first determined by other means (i.e. using the device's layout design or X-ray imaging for unknown devices).

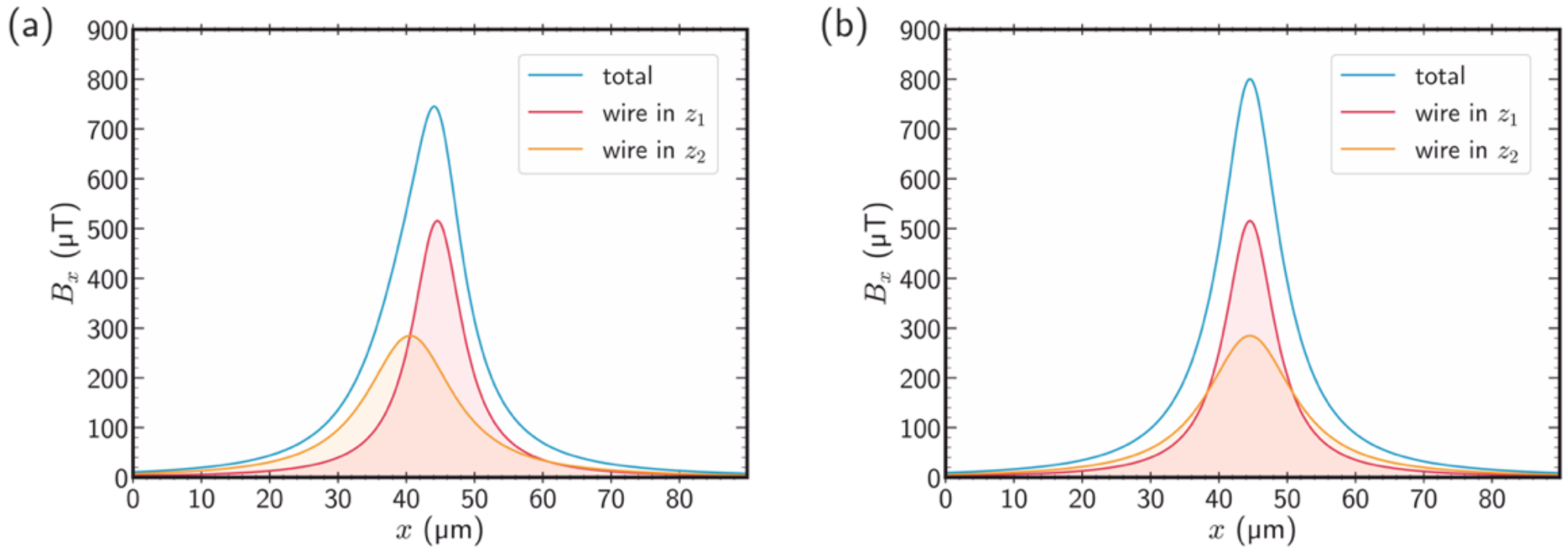

**Fig. S21. B-field profiles generated by overlapping wires. a)** $B_x$ profile generated by the partially overlapping wires, along the dashed line shown in Fig. S19. **b)** $B_x$ profile generated by the fully overlapping wires, along the dashed line shown in Fig. S20. In (a) and (b), blue lines represent the combined contribution from both wires; red lines show the contribution from the wire in the 1st AL (i.e. $z = z_1$) and orange lines show the contribution from the wire in the 2nd AL (i.e. $z = z_2$).

Finally, in the last scenario, we consider crossing wires. This time, a current $I_\alpha$ runs along with the *y*-axis in a wire in the 1st AL and the *x*-axis in a wire in the 2nd AL as shown in Fig. S22a. These currents generate orthogonal Oersted fields shown in Fig. S22, b-d. As such, the currents are unambiguously distinguishable as NV centres image the three magnetic field components separately.

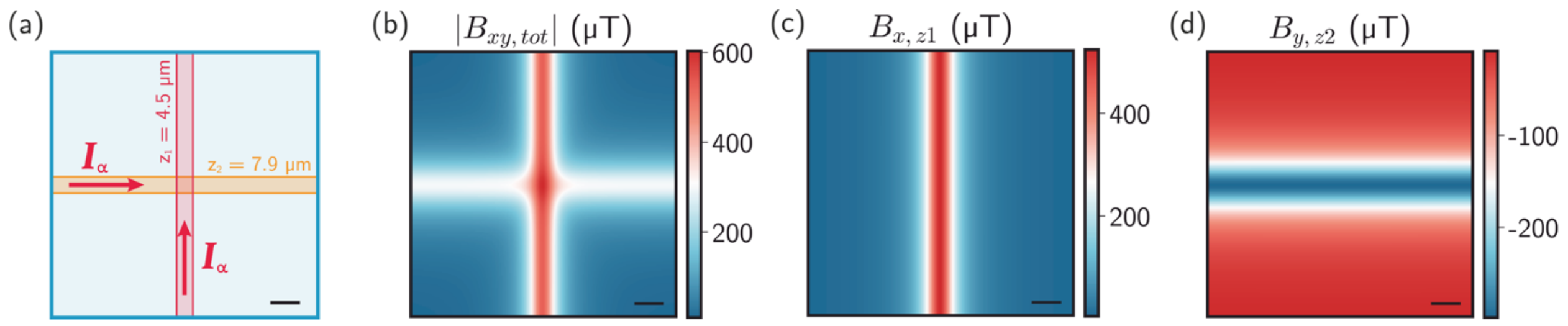

**Fig. S22. Top view of Oersted fields generated by crossing overlapping wires. a)** Schematic view of the wires. **b)** $B_x$ pattern generated by the combined contributions of the main leads in the 1st and 2nd ALs, at the sensors position. **c)** $B_x$ pattern generated by the contribution of the main lead in the 1st AL, at the sensors position. **d)** $B_x$ pattern generated by the contribution of the main lead in the 2nd AL, at the sensors position. Scale bars are 10 µm wide.

From this analysis, it is clear that NV centres provide several advantages to image currents in a multi-layered device. The unique ability to image $B_x, B_y, B_z$ separately on the nanoscale grants access to all the current density components in micro-scale devices. Additionally, it allows seeking in-depth information, especially for crossing wires. Finally, it also allows seeking these fields with the highest resolution known for the magnetic current imaging (MCI) method, as further detailed in Section S5.

## S5. Spatial resolution limitations

### S5.1. Spatial resolution of the optical imaging instrument

In imaging techniques, spatial resolution is an essential parameter to characterise as it defines the power to resolve nearby objects in real-space. For optical imaging instruments, the spatial resolution is fundamentally limited by the diffraction of light. According to Rayleigh's criterion [74], the lateral spatial resolution *xy* for our optical setup configuration is about 450 nm. Such a high spatial resolution is required here to avoid restricting the spatial resolution of the magnetic field images. For situations where a higher spatial resolution is required, techniques to image beyond the diffraction limit can be applied to the NV-based microscope [75,76].

### S5.2. Spatial resolution of the magnetic field images

One crucial aspect of any magnetic field detection technique is that resolution is ultimately limited by the sensor-source distance or the sensor size. As an NV centre is an atom-like system, the limiting factor for the spatial resolution in our case is the spacing between the sensor and the wires. Thus, it is essential to ensure the closest proximity between the NV centres and the sample. For this reason, we placed the sensor at only < 1 µm from the sample's surface. In this case, the spatial resolution of the magnetic field images is mainly limited by the architecture of the chip itself and is about a few micrometers for the magnetic field maps shown in Fig. 2.

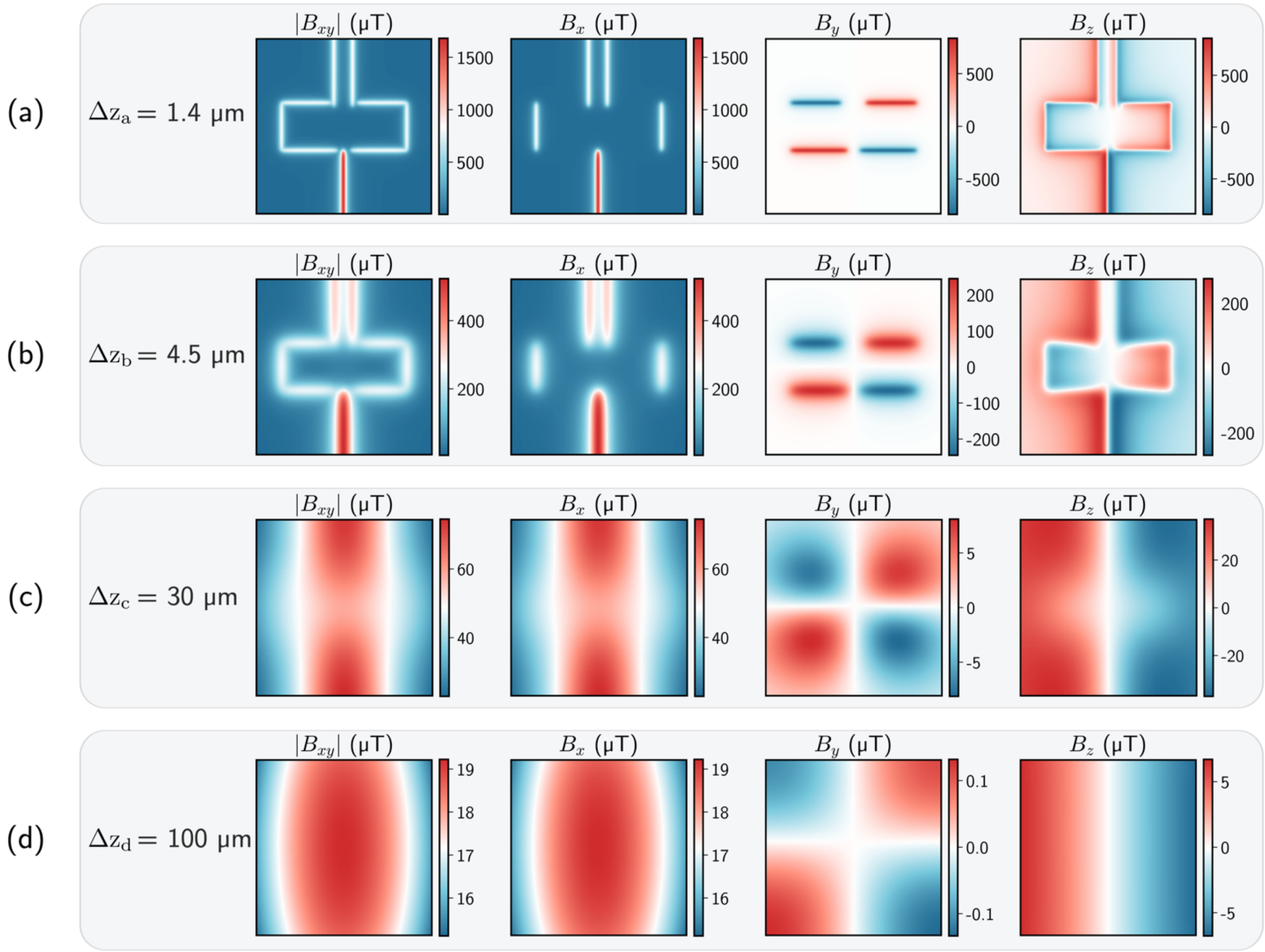

**Fig. S23. Magnetic field patterns produced by the 1st simulated AL at different distances of observation.** **a)** Magnetic patterns at $\Delta z_a = 1.4$ µm, i.e. right at the surface of the leads. **b)** Magnetic pattern at $\Delta z_b = 4.5$ µm, i.e. at the distance set by the geometry of the chip plus the 0.8 µm stand-off distance of the NV centres with the sample. **c)** Magnetic pattern at $\Delta z_c = 30$ µm. **d)** Magnetic pattern at $\Delta z_d = 100$ µm, i.e. at a standard required stand-off distance with SQUID-based imagers.

An extra distance between the sensor and the current source dramatically reduces the spatial resolution of the magnetic field images. In Fig. S23, we show the effect of several micrometres on the magnetic field pattern produced by the 1st AL.

At the surface of the leads (Fig. S23a), the magnetic features are sharp, and the pattern unambiguously reflects the geometry of the circuit. A few micrometers away, at the position of the sensors (as simulated in the main text, Fig. 4), the magnetic field slightly broadens, but no information is lost (Fig. S23b). Finally, at greater distances as shown in Fig. S23, c and d, the magnetic field features broaden such that the main pattern is lost. Additionally, the magnetic field amplitude is severely affected: nearly two orders of magnitude are lost between the magnetic field observed at $\Delta z_a = 1.4$ µm and $\Delta z_d = 100$ µm.

### S5.3. Spatial resolution of the current density images

We define the spatial resolution $\Delta x$ by the rise of the current density function across a step edge according to Sparrow's resolution criterion [77]. We determine $\Delta x$ as the distance over which the current density rises from 15 % to 85 % between its baseline and its maximum value [32,78]. The spatial resolution $\Delta x$ is about 840 nm as shown in Fig. S24 and can be further enhanced using additional regularisation methods [40] during the current reconstruction procedure.

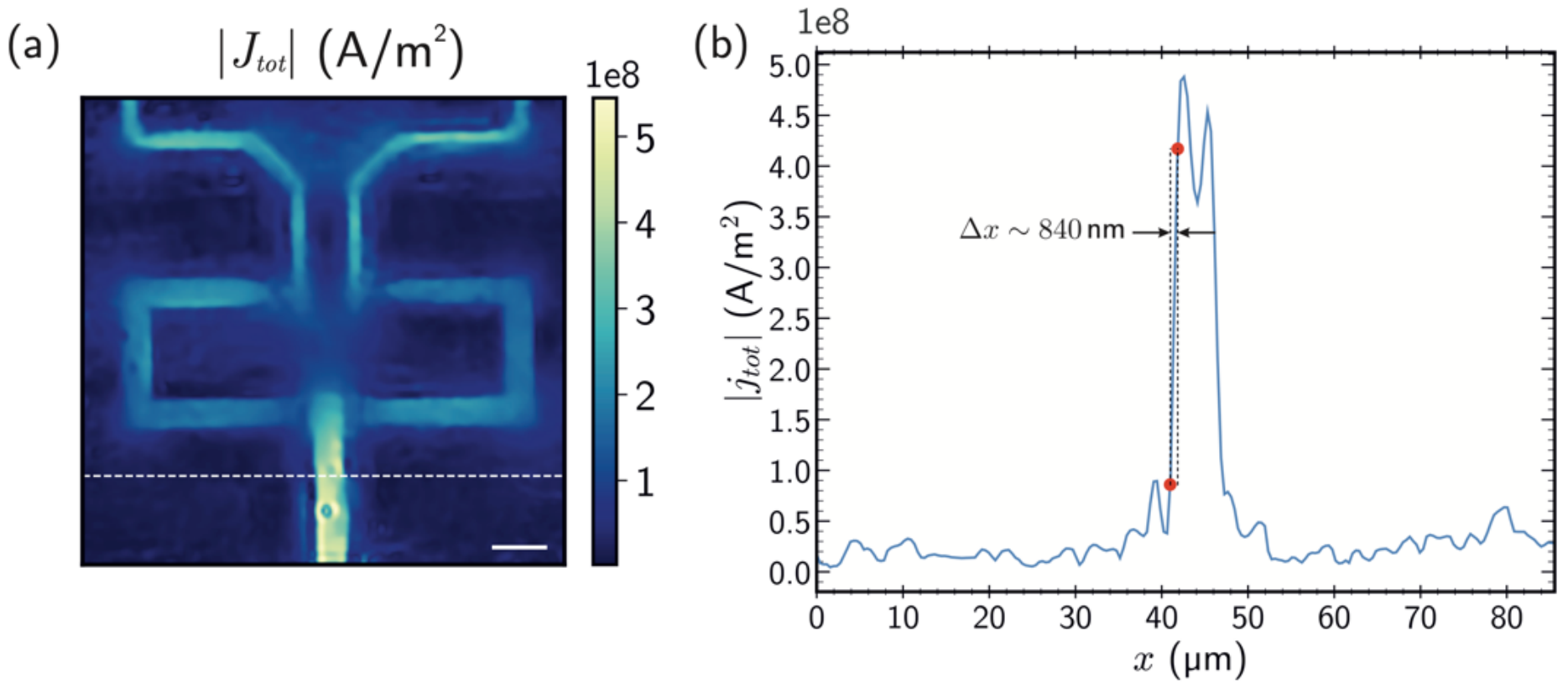

**Fig. S24. Current density spatial resolution. a)** Total current density amplitude map $|J_{\text{tot}}|$ where $|J_{\text{tot}}| = \sqrt{J_x^2 + J_y^2 + J_z^2}$. **b)** Linecut of the current density amplitude map along the white dashed line in (a). The bottom red dot represents 15 % of the value of the function's rise and the top red dot represents 85 % of the rise value. The distance between the red dots defines the spatial resolution.

## S6. Failure analysis

The main text shows how NV centres enable 3D current imaging of microscale devices and how it can serve to pint-point failure sites in modern devices. In order to define the device 1 as the operational device, we refer to the expected values of direct current in the main lead and the split branches given by the manufacturer. The values depend on an applied bias voltage and are summarised in Table S1.

We acquire the Oersted fields for each bias voltage following the procedure described in Section S1, resulting in the $B_x$ maps shown in Fig. S25. Linecuts along the *x*-axis at the position $y_A$ and $y_B$ are shown in Fig. S26. Linecuts are fitted according to Eq. (2) and the resulting current amplitudes in the main lead and the split branches are summarised in Table S2. We can see the values follow the same tendency as expected (accounting for losses due to the in-house settings such as the gold wire bonding). As such, the device investigated here is referenced as the *operational* device in the main text. Finally, the failing device is defined by comparison with the working device.

| Bias Voltage | Resulting current for main lead | Resulting current for split branches |
|---|---|---|
| no bias | $I_A =$ 10.0 mA | $I_{B1,3} =$ 5.0 mA |
| 0.5 V | $I_A =$ 8.2 mA | $I_{B1,3} =$ 4.1 mA |
| 1.0 V | $I_A =$ 10.0 mA | $I_{B1,3} =$ 5.0 mA |
| 2.0 V | $I_A =$ 13.8 mA | $I_{B1,3} =$ 6.9 mA |
| 3.0 V | $I_A =$ 17.6 mA | $I_{B1,3} =$ 8.8 mA |

**Table S1. Current depending on bias voltage, given by the manufacturer.**

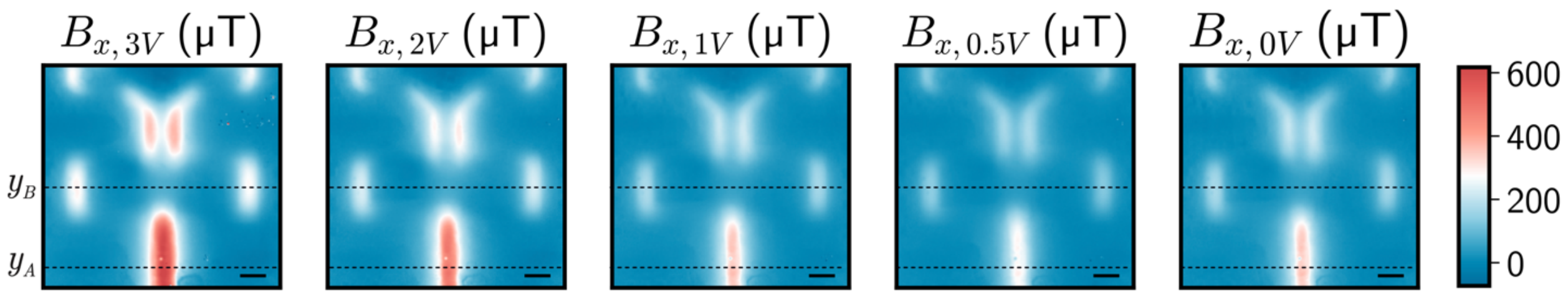

**Fig. S25. $B_x$ maps produced by various bias voltages.** From left to right: the bias voltage has been set to the following values: 3.0 V, 2.0 V, 1.0 V, 0.5 V, no bias; resulting in Oersted fields of different magnitude. Linecuts along the x-axis, at $y_A$ and $y_B$ are shown in Fig. S26. Note that the dataset for the bias voltage 2.0 V is a different dataset than shown in the main text. Scale bars are 10 μm wide.

| Bias Voltage | Resulting current for main lead | Resulting current for split branches |
|---|---|---|
| no bias | $I_A =$ 8.49(5) mA | $I_{B1} =$ 3.95(3) mA, $I_{B3} =$ 4.21(3) mA |
| 0.5 V | $I_A =$ 7.14(5) mA | $I_{B1} =$ 3.29(3) mA, $I_{B3} =$ 3.59(3) mA |
| 1.0 V | $I_A =$ 8.58(5) mA | $I_{B1} =$ 4.27(3) mA, $I_{B3} =$ 4.27(3) mA |
| 2.0 V | $I_A =$ 11.79(6) mA | $I_{B1} =$ 5.71(5) mA, $I_{B3} =$ 5.71(5)mA |
| 3.0 V | $I_A =$ 15.6(1) mA | $I_{B1} =$ 7.44(4) mA, $I_{B3} =$ 7.44(4) mA |

**Table S2. Current depending on bias voltage, extracted experimentally.**

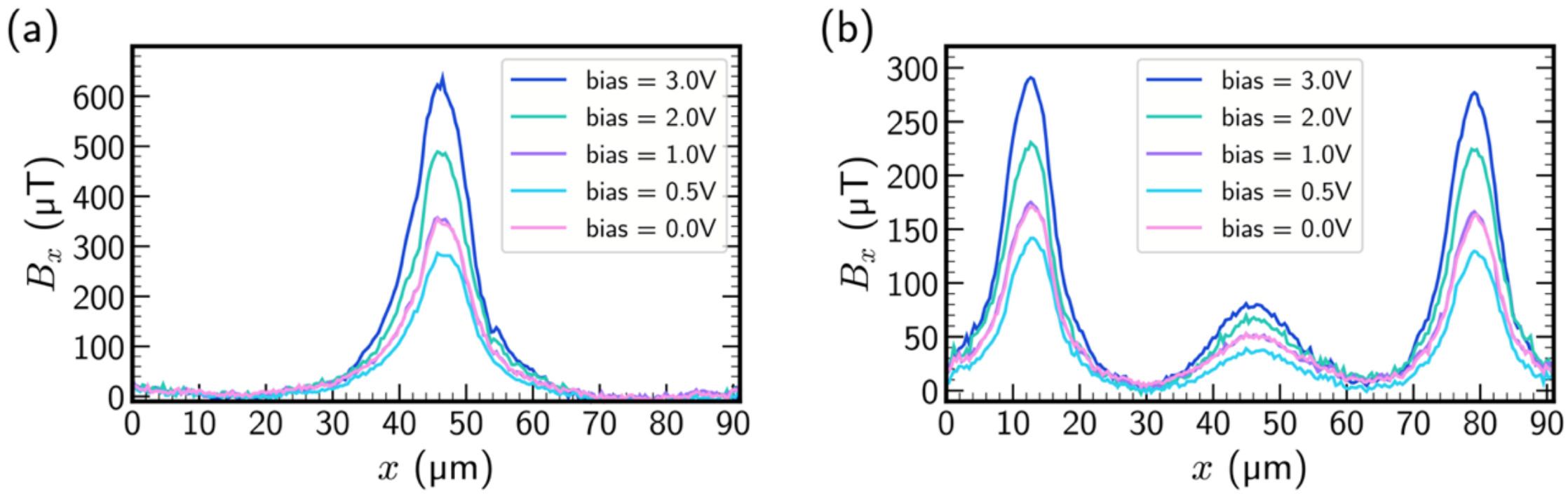

**Fig. S26. $B_x$ profiles for various bias voltages. a)** Linecut of $B_x$ along *x* at $y_A$, showing the Oersted field produced by the main lead for various bias voltages. **b)** Linecut of $B_x$ along *x* at $y_B$, showing the Oersted field produced by the split branches for various bias voltages.

## References

[22] Doherty, M. W., Manson, N. B., Delaney, P., Jelezko, F., Wrachtrup, J. & Hollenberg, L. C. L. The nitrogen-vacancy colour centre in diamond. *Phys Rep* **528,** 1–45 (2013).

[24] Steinert, S., Dolde, F., Neumann, P., Aird, A., Naydenov, B., Balasubramanian, G., Jelezko, F. & Wrachtrup, J. High sensitivity magnetic imaging using an array of spins in diamond. *Review of Scientific Instruments* **81,** 043705 (2010).

[32] Chang, K., Eichler, A., Rhensius, J., Lorenzelli, L. & Degen, C. L. Nanoscale Imaging of Current Density with a Single-Spin Magnetometer. *Nano Lett* **17,** 2367–2373 (2017).

[37] Böck, J., Aufinger, K., Boguth, S., Dahl, C., Knapp, H., Liebl, W., Manger, D., Meister, T. F., Pribil, A., Wursthorn, J., Lachner, R., Heinemann, B., Rücker, H., Fox, A., Barth, R., Fischer, G., Marschmeyer, S., Schmidt, D., Trusch, A. & Wipf, C. SiGe HBT and BiCMOS process integration optimization within the DOTSEVEN project. *Proceedings of the IEEE Bipolar/BiCMOS Circuits and Technology Meeting* 121–124 (2015).

[39] Roth, B. J., Sepulveda, N. G. & Wikswo, J. P. Using a magnetometer to image a two-dimensional current distribution. *J Appl Phys* **65,** 361–372 (1989).

[40] Meltzer, A. Y., Levin, E. & Zeldov, E. Direct Reconstruction of Two-Dimensional Currents in Thin Films from Magnetic-Field Measurements. *Phys Rev Appl* **8,** 064030 (2017).

[60] Mistrik, J., Kasap, S., Ruda, H. E., Koughia, C. & Singh, J. Optical properties of electronic materials: fundamentals and characterization. *Springer Handbooks* 1 (2017).

[61] Zeleznik, F. J. Quasi-Newton Methods for Nonlinear Equations. *Journal of the Association for computing Machinery* **15,** 265–271 (1968).

[62] Levenberg, K. A method for the solution of certain non-linear problems in least squares. *Q Appl Math* **2,** 164–168 (1944).

[63] Marquardt, D. W. An Algorithm for Least-Squares Estimation of Nonlinear Parameters. *Journal of the Society for Industrial and Applied Mathematics* **11,** 431–441 (1963).

[64] Dréau, A., Lesik, M., Rondin, L., Spinicelli, P., Arcizet, O., Roch, J. F. & Jacques, V. Avoiding power broadening in optically detected magnetic resonance of single NV defects for enhanced dc magnetic field sensitivity. *Physical Review B* **84,** 195204 (2011).

[65] Udvarhelyi, P., Shkolnikov, V. O., Gali, A., Burkard, G. & Pályi, A. Spin-strain interaction in nitrogen-vacancy centers in diamond. *Phys Rev B* **98,** 075201 (2018).

[66] Acosta, V. M., Bauch, E., Ledbetter, M. P., Waxman, A., Bouchard, L. S. & Budker, D. Temperature dependence of the nitrogen-vacancy magnetic resonance in diamond. *Phys Rev Lett* **104,** 070801 (2010).

[67] Doherty, M. W., Acosta, V. M., Jarmola, A., Barson, M. S. J., Manson, N. B., Budker, D. & Hollenberg, L. C. L. Temperature shifts of the resonances of the NV-center in diamond. *Phys Rev B Condens Matter Mater Phys* **90,** 041201 (2014).

[68] Dolde, F., Fedder, H., Doherty, M. W., Nöbauer, T., Rempp, F., Balasubramanian, G., Wolf, T., Reinhard, F., Hollenberg, L. C. L., Jelezko, F. & Wrachtrup, J. Electric-field sensing using single diamond spins. *Nat Phys* **7,** 459–463 (2011).
[69] Broadway, D. A., Dontschuk, N., Tsai, A., Lillie, S. E., Lew, C. T. K., McCallum, J. C., Johnson, B. C., Doherty, M. W., Stacey, A., Hollenberg, L. C. L. & Tetienne, J. P. Spatial mapping of band bending in semiconductor devices using in situ quantum sensors. *Nat Electron* **1,** 502–507 (2018).
[70] Schlussel, Y., Lenz, T., Rohner, D., Bar-Haim, Y., Bougas, L., Groswasser, D., Kieschnick, M., Rozenberg, E., Thiel, L., Waxman, A., Meijer, J., Maletinsky, P., Budker, D. & Folman, R. Wide-Field Imaging of Superconductor Vortices with Electron Spins in Diamond. *Phys Rev Appl* **10,** 034032 (2018).
[71] Goodall, C. R. 13 Computation using the QR decomposition. *Handbook of Statistics* **9,** 467–508 (1993).
[72] Chambolle, A. An Algorithm for Total Variation Minimization and Applications. *J Math Imaging Vis* **20,** 89–97 (2004).
[73] Broadway, D. A., Lillie, S. E., Scholten, S. C., Rohner, D., Dontschuk, N., Maletinsky, P., Tetienne, J.-P. & Hollenberg, L. C. L. Improved Current Density and Magnetization Reconstruction Through Vector Magnetic Field Measurements. *Phys Rev Appl* **14,** 024076 (2020).
[74] Rayleigh, J. W. S. XV. On the theory of optical images, with special reference to the microscope. *The London, Edinburgh, and Dublin Philosophical Magazine and Journal of Science* **42,** 167–195 (1896).
[75] Beyond the diffraction limit. *Nat Photonics* **3,** 361 (2009).
[76] Pfender, M., Aslam, N., Waldherr, G., Neumann, P. & Wrachtrup, J. Single-spin stochastic optical reconstruction microscopy. *Proceedings of the National Academy of Sciences* **111,** 14669–14674 (2014).
[77] Sparrow, C. M. On Spectroscopic Resolving Power. *Astrophys J* **44,** 76 (1916).
[78] Yu, T., Liu, H. & Cai, W. On the quantification of spatial resolution for three-dimensional computed tomography of chemiluminescence. *Opt Express* **25,** 24093 (2017).